\pdfoutput=1
\documentclass[preprint,3p]{elsarticle}
\journal{Physica D}

\usepackage{graphicx}
\usepackage{epstopdf}
\usepackage{epsfig}
\DeclareGraphicsExtensions{-eps-converted-to.pdf}
\usepackage{pdfpages}

\usepackage{amssymb}
\usepackage{empheq}
\usepackage{cases}
\usepackage{amsthm,amsmath}
\usepackage{caption,lipsum}
\newcommand{\eps}{\varepsilon}

\newcommand{\bR}{\mathbb{R}}
\newcommand{\bC}{\mathbb{C}}
\newcommand{\bv}{\mathbf{v}}

\newcommand{\fe}{\mathrm{e}}
\newcommand{\bk}{\textbf{k}}

\newcommand{\bx}{\mathbf{x}}
\newcommand{\by}{\mathbf{y}}
\newcommand{\ud}{\mathrm{d}}
\newcommand{\sech}{\mathrm{sech}}

\newtheorem{remark}{Remark}[section]

\newcommand{\be}{\begin{equation}}
\newcommand{\ee}{\end{equation}}
\newcommand{\ba}{\begin{array}}
\newcommand{\ea}{\end{array}}
\newcommand{\bea}{\begin{eqnarray}}
\newcommand{\eea}{\end{eqnarray}}
\newcommand{\beas}{\begin{eqnarray*}}
\newcommand{\eeas}{\end{eqnarray*}}
\input psfig.sty

\begin{document}

\begin{frontmatter}

\title{A modulation equations approach for numerically solving the moving soliton and radiation solutions of NLS}

\author[US]{Avy Soffer}
 \ead{soffer@math.rutgers.edu}

\address[US]{Department of Mathematics, Rutgers University, New Jersey, 08854, USA}
\address[FA]{IRMAR, Universit\'{e} de Rennes 1, Rennes, 35042, France}

\author[FA]{Xiaofei Zhao\corref{5}}
\ead{zhxfnus@gmail.com}

\cortext[5]{Corresponding author. Tel : +33 223235385; fax : +33 223235464.}

\begin{abstract}
Based on our previous work for solving the nonlinear Schr\"{o}dinger equation with multichannel dynamics that is given by a localized standing wave and radiation, in this work we deal with the multichannel solution which consists of a moving soliton and radiation.
We apply the modulation theory to give a system of ODEs coupled to the radiation term for describing the solution, which is valid for all times. The modulation equations are solved accurately by the proposed numerical method. The soliton and radiation are captured separately in the computation, and they are solved on the translated domain that is moving with them. Thus for a fixed finite physical domain in the lab frame, the multichannel solution can pass through the boundary naturally, which can not be done by imposing any existing boundary conditions.
We comment on the differences of this method from the collective coordinates.
\end{abstract}


\end{frontmatter}

\section{Introduction}
\label{sec: intr}
We consider the following nonlinear Schr\"{o}dinger (NLS) equation in $d$ dimensions ($d=1,2,3$)
 \begin{align}
&i\partial_t u(\mathbf{x},t)+\Delta u(\mathbf{x},t)+\beta(|u(\mathbf{x},t)|^{2})u(\mathbf{x},t)=0,\quad \mathbf{x}\in\bR^d,\ t>0,\label{nls}\\
&u(\mathbf{x},0)=u_0(\mathbf{x}),\quad \mathbf{x}\in\bR^d,
\end{align}
where $u_0(\mathbf{x})$ is the given initial data and $\beta(\cdot):\bR\to\bR$ is a smooth nonlinear function. It is well-known that the mass of the system $M$ and the Hamiltonian (energy) of the system $H$ are conserved, i.e.
\begin{align}
M(t)&:=\int_{\bR^d}\left|u(\mathbf{x},t)\right|^2\ud \mathbf{x}\equiv M(0),\\
H(t)&:=\int_{\bR^d}\left[\left|\nabla u(\mathbf{x},t)\right|^2-F\left(\left|u(\mathbf{x},t)\right|^{2}\right)
\right]\ud \mathbf{x}\equiv H(0),\label{E&M}
\end{align} with $F'(\rho)=\beta(\rho)$.
We study the multichannel dynamics in (\ref{nls}).
In fact, many physical systems such as the particles or matter wave dynamics in the quantum mechanics and the nonlinear optics \cite{Malomed,BBB2,Morales3,Morales4,Nishida,374}, involve the dynamics of the multichannel solution which mean the solution of the system asymptotically is given by a linear combination of a localized (in space), periodic (in time) wave (solitary or standing wave) and a dispersive part \cite{Soffer1}. The dispersive part is usually referred as radiation in the literature.
The multichannel solutions also widely exist in many other conservative nonlinear dispersive and wave equations besides the NLS equation \cite{Soffer1}, and they have been found of importance and useful in both theoretical analysis and applications.

In this paper, we focus on the study of the multichannel solution of a moving soliton and radiation in the NLS equation (\ref{nls}), i.e. the solution $u(\bx,t)$ of (\ref{nls}) is given by
\begin{equation}\label{multichannel}
u(\bx,t)=\Phi_{soliton}(\bx,t)+\Phi_{radiation}(\bx,t),
\end{equation}
where $\Phi_{radiation}(\bx,t)$ denotes the dispersive wave and the soliton $\Phi_{soliton}(\bx,t)$ has form
\begin{equation*}
\Phi_{soliton}(\bx,t)=\fe^{i\theta(\bx,t)}\phi_\omega(\bx-\bv t-D),\qquad \bx\in\bR^d,\ t>0,
\end{equation*}
with $\bv\in\bR^d$ the velocity, $D\in\bR^d$ the shift in space, $\theta(\bx,t)\in\bR$ the phase and $\phi_\omega(\bx)>0$ the eigenfunction of $\omega>0$, i.e.
\begin{equation*}
\omega\phi_\omega(\bx)=\Delta\phi_\omega(\bx)+\beta(\phi_\omega^2(\bx))\phi_\omega(\bx),\qquad \bx\in\bR^d.
\end{equation*}
When the system is free from radiation, the phase function is given by
\begin{equation}\label{free}
\theta(\bx,t)=\frac{\bv\cdot\bx}{2}-\frac{|\bv|^2t}{4}+\omega t+\gamma,
\end{equation}
 where $\gamma\in\bR$ is the shift in phase. In the presence of radiation, it then becomes a function in general and the parameters $\bv,D,\omega$ become functions of time.

In the studies of the multichannel dynamics, a very popular method in the physics community is the collective coordinates method. The collective coordinates method usually begins with a guess of the soliton of the system and drops the radiation, which will result in some ODEs to describe the physical system approximately. Thus it often applies when the soliton part is the main interest and the radiation is small. For a detailed review of this method, we refer the readers to \cite{374} and the references therein. However as pointed out in \cite{Zhao}, this method can only provide a good approximation of the soliton for short time. For the long time dynamics or large initial radiation, the collective coordinates method will fail. Another major approach to study the multichannel dynamics is to numerically solving the governing equations by truncating the whole space problem onto a finite domain and then imposing some suitable boundary conditions. The imposed boundary condition for computation is the key to maintain the true physics. Classical Dirichlet or
Neumann boundaries introduce spurious reflections, while periodic boundaries allow outgoing waves to wrap around the computational domain. For the wave equations, the Dirichlet-to-Neumann method that attempts to use the exact solution as a boundary condition, has been established and works very well \cite{Bayliss,Bayliss2,Engquist2,Engquist3,Givoli,Hagstrom}. However it is fraught with problems for dispersive equations \cite{Lubich,Schadel,Szeftel} and only limited progress has been made so far. People then used a dissipative term which is localized on a buffer region to dissipates outgoing waves, but this method also dissipates incoming waves located near the boundary which is spurious \cite{Berenger,Baer}. By decomposing the solution into a family of coherent states, a phase space filter method has been proposed to design an open boundary for the NLS equation in \cite{SStu,Stucchio}, but this method fails to filter waves with wavelength longer than the buffer region, which is also the problem shared with most absorbing boundary conditions. In \cite{Stucchio}, by combining the phase space filter method to a spectral technique that resolves waves of both long wavelength and short wavelength, a multiscale method has been considered to filter outgoing waves regardless of the frequency.

To study the multichannel solution (\ref{multichannel}) in (\ref{nls}), we are going to apply the modulation theory, which has been established in \cite{Cuccagna} for studying the stabilization of solution in the NLS equation from the asymptotical point of view. Here the modulation equations for governing the multichannel solution are derived exactly. They are a system of ODEs coupled to a dispersive equation for the radiation and they can exactly describe the dynamics of the solitary wave and the dispersive part
separately at the same time, which are valid for all times. Thanks to our recent work in \cite{Zhao} where we studied the multichannel dynamics of a standing wave and radiation, an efficient and accurate numerical method is proposed to solve the modulation equations. The modulation equations are indeed solved numerically on the Lagrangian domain that is moving with the soliton. Thus, for any fixed finite domain in the lab frame, the multichannel solution can approach and pass through the boundaries naturally, which can not be done by any boundary techniques as we discussed.
Numerical explorations are provided in the end towards better understanding of the multichannel dynamics. Although the work here is done for the NLS equation, it is believed that the approach could work for other dispersive equations.

The rest of the paper is organized as follows. In Section \ref{sec: review}, we shall derive the modulation equations with a brief review. In Section \ref{sec: eigen pro}, we shall propose an algorithm for the modulation equations. Numerical studies and results are given in Section \ref{sec: result} followed by some discussions.

\section{Modulation equations}\label{sec: review}
In this section, we shall formally derive the full modulation equations of the NLS equation for describing the multichannel solution (\ref{multichannel}) by following the sketch used in \cite{Cuccagna} and review some related mathematical theories for the readers' convenience. Comments on the differences of this method from the well-known method of collective coordinates are given in the end.

\subsection{Formal derivation}
Based on the fact of the free soliton (\ref{free})
in the NLS equation (\ref{nls}) and the physical observation that the qualitative behavior of the linear Schr\"{o}dinger equation should not change that much in
response to a small nonlinear and Hamiltonian perturbation in the dynamics, i.e.
we should still see a localized part which decouples after a long time from the
dispersive part, thus we take the ansatz of the solution of (\ref{nls}) \cite{Cuccagna} as
\begin{align}
&u(\mathbf{x},t)=\fe^{i\theta(\bx,t)}\left[\phi_{\omega(t)}\left(\by(\mathbf{x},t)\right)+R\left(\by(\mathbf{x},t),t\right)\right],\quad\mathbf{x}\in\bR^d,\ \,t\geq0,\tag{A1}\label{A1}\\
&\theta(\bx,t)=\frac{1}{2}\bv\cdot\bx-\frac{1}{4}\int_0^t|\bv(s)|^2\ud s+\int_0^t\omega(s)\ud s+\gamma(t),\tag{A2}\\
&\by(\bx,t)=\bx-\int_0^t\bv(s)\ud s-D(t),\tag{A3}\label{A3}
\end{align}
with initial conditions
\begin{align}
&u_0(\bx)=\fe^{i\left(\frac{\bv_0\cdot\bx}{2}+\gamma_0\right)}\left[\phi_{\omega_0}(\bx-D_0)+R_0(\bx-D_0)\right],\nonumber\\
&\bv(0)=\bv_0,\quad \omega(0)=\omega_0,\quad \gamma(0)=\gamma_0,\quad D(0)=D_0,\quad R(\bx,0)=R_0(\bx).\label{modulation: ini}
\end{align}
Here, $\phi_\omega(\bx)$ is the nonlinear bound state of the time-independent NLS equation of (\ref{nls}) with eigenvalue $\omega$, i.e.
\begin{align}
&\omega\phi_\omega(\bx)=\Delta\phi_\omega(\bx)+\beta(\phi_\omega^2(\bx))\phi_\omega(\bx),\qquad \bx\in\bR^d,\label{eigenfunction}\\
& \phi_\omega(\bx)\in H^2(\bR^d),\qquad \phi_\omega>0.\nonumber
\end{align}
$\by\in\bR^d$ is interpreted as the translated space variable or the Lagrangian domain which is moving with the solution with velocity $\bv$. $\phi_\omega(\by)$ is the moving soliton in the multichannel solution and $R(\by,t)$ is the dispersive wave. Under some conditions on the nonlinearity $\beta(\cdot)$, for $\omega>0$, the eigenvalue problem (\ref{eigenfunction}) has a unique radially symmetric solution exponential decaying at far field \cite{Cuccagna,Soffer1}.

Plugging the ansatz (\ref{A1})-(\ref{A3}) into the NLS equation (\ref{nls}), we obtain
\begin{align}
&i \partial_tR+\left(\partial_{\by\by}-i\dot{D}\cdot\partial_\by \right) R-\left(\omega+\dot{\gamma}+\frac{\dot{\bv}\cdot\bx}{2}\right)R-i\dot{D}\cdot\partial_\by\phi_{\omega}+i\dot{\omega}\partial_\omega\phi_\omega\label{R eq0}\\
&+\partial_{\by\by}
\phi_\omega-\left(\omega+\dot{\gamma}+\frac{\dot{\bv}\cdot\bx}{2}\right)\phi_\omega+\beta\left(|\phi_\omega+R|^2\right)(\phi_\omega+R)=0,\nonumber\ \bx\in\bR^d,\ t>0,
\end{align}
where $R=R(\by(\bx,t),t)$, $\phi_\omega=\phi_{\omega(t)}(\by(\bx,t))$, $\partial_\omega\phi_\omega=\partial_\omega\phi_{\omega(t)}(\by(\bx,t))$, $\partial_\by$ and $\partial_{\by\by}$ denote the gradient and Laplacian operators with respect to the variable $\by$ respectively, and the function $\partial_\omega\phi_\omega(\bx)$ is given by
\begin{align}\label{Domega}
\omega\partial_\omega\phi_{\omega}(\bx)=\left[\Delta+2\beta'(\phi_\omega^2(\bx))\phi_{\omega}^2(\bx)
+\beta(\phi_\omega^2(\bx))\right]\partial_\omega\phi_\omega(\bx)-\phi_\omega(\bx),
\quad \bx\in\bR^d.
\end{align}
Since $\phi_\omega(\bx)$ is even, so it is clear that $\partial_\omega\phi_{\omega}(\bx)$ is also an even function.
By introducing a new parameter $\xi(t)$ with arbitrary initial value $\xi(0)=\xi_0\in\bR$ and
\begin{align}\label{xi def}
\dot{\xi}(t)=\dot{\gamma}(t)+\frac{1}{2}\dot{\bv}(t)\cdot\int_0^t\bv(s)\ud s+\frac{1}{2}\dot{\bv}(t)\cdot D(t), \quad t>0,
\end{align}
which indicates that
$$\dot{\gamma}+\frac{\dot{\bv}\cdot\bx}{2}=\dot{\xi}+\frac{\dot{\bv}\cdot\by}{2},$$
 the explicit dependence on the variable $\bx$ in
(\ref{R eq0}) can be removed, and the equation can be interpreted as imposed on the $\by$-domain, i.e. the translated Lagrangian domain in terms of $\by$. By further decompose the interaction term into the linear part of $R$ and the nonlinear part denoted as $N=N(\by,t)$, i.e.
\begin{align}
\beta\left(|\phi_\omega+R|^2\right)(\phi_\omega+R)=\beta\left(\phi_\omega^2\right)\phi_\omega+\beta\left(\phi_\omega^2\right)R+
\beta'\left(\phi_\omega^2\right)\phi_\omega^2\cdot(R+\overline{R})+N,
\end{align}
where here and after $\overline{z}$ denotes the complex conjugate of a complex number $z$. The nonlinear part $N=O(|R|^2)$, as $|R|\to0$. For the cubic nonlinearity case, i.e. $\beta(\rho)=\lambda\rho,\ \lambda\in\bR$, we have
$$N=\lambda\phi_\omega\left(R^2+|R|^2\right)+\lambda(\phi_\omega+R)|R|^2.$$
Then by also noting (\ref{eigenfunction}), the equation (\ref{R eq0}) can be rewritten as
\begin{align}
&i \partial_tR(\by,t)+\left(\partial_{\by\by}-i\dot{D}\cdot\partial_\by \right) R(\by,t)-\left(\omega+\dot{\xi}+\frac{\dot{\bv}\cdot\by}{2}\right)R(\by,t)
\label{R eq1}\\
&+\left[\beta(\phi_\omega^2(\by))+\beta'(\phi_\omega^2(\by))\phi_\omega^2(\by)\right]R(\by,t)
+\beta'(\phi_\omega^2(\by))\phi_\omega^2(\by)\overline{R}(\by,t)+N(\by,t)\nonumber\\
&-i\dot{D}\cdot\partial_\by\phi_{\omega}(\by)+i\dot{\omega}\partial_\omega\phi_\omega(\by)-\left(\dot{\xi}+\frac{\dot{\bv}\cdot\by}{2}\right)\phi_\omega(\by)
=0,\nonumber\quad \by\in\bR^d,\ t>0.
\end{align}
As a convention \cite{Cuccagna,Schlag}, in the following, we interpret $R=R_1+iR_2\in\bC$ which is originally a complex-valued scalar function, as a two-dimensional column vector function and so do other terms in (\ref{R eq1}), i.e.
$$R(\by,t)=\left(\begin{split}&R_1(\by,t)\\&R_2(\by,t)\end{split}\right),\quad
\phi_\omega(\by)=\left(\begin{split}&\phi_\omega(\by)\\&\quad 0\end{split}\right),\quad
N(\by,t)=\left(\begin{split}&N_1(\by,t)\\&N_2(\by,t)\end{split}\right),$$
where $R_1$ and $R_2$ denote the real part and imaginary part of $R$ respectively and $N_1$ and $N_2$ denote the real part and imaginary part of $N$ respectively.
Introduce two two-by-two matrixes
$$J=\left(\begin{split}0\quad &1\\-1 \quad &0\end{split}\right),\qquad H(t)=\left(\begin{split}&L_+\ \ 0 \\ &0 \quad\ L_-\end{split}\right),$$
with time dependent operators
$$L_+:=-\partial_{\by\by}+\omega-\beta(\phi_\omega^2)-2\beta'(\phi_\omega^2)\phi_\omega^2,\quad L_-:=-\partial_{\by\by}+\omega-\beta(\phi_\omega^2).$$
Here the matrix $J$ satisfying $J^2=-Id$ is introduced as the matrix representation of the imaginary unit $-i$. Then (\ref{R eq1}) can be written into a real-valued vector form as
\begin{align}
\partial_tR(\by,t)=&JH(t)R(\by,t)+\left(\dot{D}\cdot\partial_\by \right) R(\by,t)+\left(\dot{\xi}+\frac{\dot{\bv}\cdot\by}{2}\right)JR(\by,t)-JN(\by,t)
\nonumber\\
&+\left(\dot{\xi}+\frac{\dot{\bv}\cdot\by}{2}\right)J\phi_\omega(\by)+\left(\dot{D}\cdot\partial_\by\right)\phi_{\omega}(\by)
-\dot{\omega}\partial_\omega\phi_\omega(\by),\quad \by\in\bR^d,\ t>0.\label{R matrix}
\end{align}
To make (\ref{R matrix}) solvable, we need to impose some extra conditions on the soliton and the radiation. We assume
$R\in\{\phi_\omega,J\partial_\omega\phi_\omega,\by\phi_\omega, J\partial_\by\phi_\omega\}^\bot$, which is known as the orthogonality conditions in the literature \cite{Comech,Cuccagna,Marzuola,Weistein}, and here the inner product of two vectors are interpreted as usual. In details, the orthogonality conditions read
\begin{equation}\label{orthog}
\left\{\begin{split}
&\left<R_1(t,\cdot),\,\phi_{\omega(t)}\right>=0,\qquad\quad \left<R_2(t,\cdot),\,\partial_\omega\phi_{\omega(t)}\right>=0,\\%\left<\partial_tR_1(t,\cdot),\,\phi_{\omega(t)}\right>=0,\\
%&\qquad \left<\partial_tR_2(t,\cdot),\,\partial_\omega\phi_{\omega(t)}\right>=0,\\
&\left<R_1(t,\cdot),\,\by\phi_{\omega(t)}\right>=\mathbf{0},\qquad \ \left<R_2(t,\cdot),\,\partial_\by \phi_{\omega(t)}\right>=\mathbf{0},%\left<\partial_tR_1(t,\cdot),\,\by\phi_{\omega(t)}\right>=\mathbf{0},\\
%&\qquad \left<\partial_tR_2(t,\cdot),\,\partial_\by\phi_{\omega(t)}\right>=\mathbf{0},
\end{split}\right.\quad t\geq0,
\end{equation}
where $<f,g>:=\int_{\bR^d}f(\bx)\cdot g(\bx)\ud\bx$ for two real-valued scalar or vector functions $f,g\in L^2(\bR^d)$.

Now we firstly take the inner product of (\ref{R matrix}) with $\phi_\omega(\by)$ on both sides. By noting that
$<\phi_\omega,\,(\dot{D}\cdot\partial_\by)\phi_\omega>=\dot{D}\cdot<\phi_\omega,\,\partial_\by\phi_\omega>=0$, we get
\begin{align}\label{omega}
\dot{\omega}\left<\phi_\omega-R_1,\,\partial_\omega\phi_\omega\right>=
\frac{\dot{\bv}}{2}\left<\phi_\omega,\,\by R_2\right>+
\dot{D}\cdot\left<\phi_\omega,\,\partial_\by R_1\right>+\dot{\xi}\left<\phi_\omega,\,R_2\right>-\left<\phi_\omega,\,N_2\right>,\quad t>0.
\end{align}
Secondly, noting the fact that
$$\left<\partial_\omega\phi_\omega,\,L_+ R_1\right>=-\left<\phi_\omega,\,R_1\right>=0,$$
which is indicated by integration by parts and (\ref{Domega}),
then by taking the inner product of (\ref{R matrix}) with $J\partial_\omega\phi_\omega(\by)$ on both sides, we get
\begin{align}
\dot{\xi}\left(\left<\partial_\omega\phi_\omega,\,R_1\right>
+\left<\phi_\omega,\,\partial_\omega\phi_\omega\right>\right)=&\dot{\omega}\left<\partial_\omega^2\phi_\omega,\,R_2\right>
-\frac{\dot{\bv}}{2}\cdot\left(\left<\partial_\omega\phi_\omega,\,\by R_1\right>
+\left<\partial_\omega\phi_\omega,\,\by \phi_\omega\right>\right)\nonumber\\
&+\dot{D}\cdot\left<\partial_\omega\phi_\omega,\,\partial_\by R_2\right>+\left<\partial_\omega\phi_\omega,\,N_1\right>,\quad t>0.\label{xi}
\end{align}
Here $\partial_\omega^2\phi_\omega$ is given by
\begin{equation}\label{d2wphi}\left[\omega-\Delta-2\beta'(\phi_\omega^2)\phi_\omega^2-\beta(\phi_\omega^2)\right]\partial_\omega^2\phi_\omega
=\left[-2+4\beta''(\phi_\omega^2)\phi_\omega^3\partial_\omega\phi_\omega+6\beta'(\phi_\omega^2)\phi_\omega
\partial_\omega\phi_\omega\right]\partial_\omega\phi_\omega.
\end{equation}
Thirdly, by taking the inner product of (\ref{R matrix}) with $J\partial_\by\phi_\omega(\by)$ on both sides and noting that $<\partial_\by\phi_\omega,\by\phi
_\omega>=-\frac{1}{2}\|\phi_\omega\|_{L^2}^2$,
we get
\begin{align}
\dot{\bv}\left(\frac{1}{2}\left<\partial_\by\phi_\omega,\,\by R_1\right>
-\frac{1}{4}\|\phi_\omega\|_{L^2}^2\right)=&-\dot{\omega}\left<\partial_\omega\phi_\omega,\,\partial_\by R_2\right>+ \dot{D}\left<\partial_\by\phi_\omega,\,\partial_\by R_2\right>
-\dot{\xi}\left<\partial_\by\phi_\omega,\, R_1\right>\nonumber\\
&-\left<\partial_\by\phi_\omega,\,L_+R_1\right>+\left<\partial_\by\phi_\omega,\,N_1\right>,\quad t>0.\label{v}
\end{align}
At last, by taking the inner product of (\ref{R matrix}) with $\by\phi_\omega(\by)$ on both sides,
we can get
\begin{align}
\dot{D}\left(\left<\by\phi_\omega,\, \partial_\by R_1\right>
-\frac{1}{2}\|\phi_\omega\|_{L^2}^2\right)=&\dot{\omega}\left<\phi_\omega-R_1,\,\by\partial_\omega \phi_\omega\right> -\frac{\dot{\bv}}{2}\left<\by\phi_\omega,\,\by R_2\right>
-\dot{\xi}\left<\by\phi_\omega,\, R_2\right>\nonumber\\
&-\left<\by\phi_\omega,\,L_-R_2\right>+\left<\by\phi_\omega,\,N_2\right>,\quad t>0.\label{D}
\end{align}
Finally, combing (\ref{omega}), (\ref{xi}), (\ref{v}) and (\ref{D}), defining
$$\sigma(t):=(\omega(t),\bv(t),D(t),\xi(t))^T\in\bR^{2d+2},\quad t\geq0,$$
a matrix $A(t):=$
$$\small \displaystyle \left(\begin{matrix}
<\phi_\omega-R_1,\partial_\omega\phi_\omega> & -\frac{1}{2}<\phi_\omega,\by R_2> & -<\phi_\omega,\partial_\by R_1> &-<\phi_\omega,R_2>\\
-<\partial_\omega^2\phi_\omega,R_2>& \frac{1}{2}<\partial_\omega\phi_\omega,\by (R_1+\phi_\omega)> & -<\partial_\omega\phi_\omega,\partial_\by R_2> & <\partial_\omega\phi_\omega,R_1+\phi_\omega>\\
<\partial_\omega\phi_\omega,\partial_\by R_2>& \frac{1}{4}(2<\partial_\by\phi_\omega,\by R_1>-\|\phi_\omega\|_{L^2}^2 )&-<\partial_\by\phi_\omega,\partial_\by R_2>&<\partial_\by\phi_\omega,R_1>\\
<R_1-\phi_\omega,\by\partial_\omega\phi_\omega)&\frac{1}{2}<\by\phi_\omega,\by R_2>&<\by\phi_\omega,\partial_\by R_1>-\frac{1}{2}\|\phi_\omega\|_{L^2}^2
&<\by\phi_\omega,R_2>
\end{matrix}\right),$$
for $t>0$ and a column vector
$$F(t):=\left(\begin{matrix}- <\phi_\omega,\,N_2>\\ <\partial_\omega\phi_\omega,\,N_1>\\ <\partial_\by\phi_\omega,\,N_1-L_+R_1>\\
<\by\phi_\omega,\, N_2-L_-R_2>\end{matrix}\right),\quad t> 0,$$
then together with (\ref{R matrix}), we get the full \emph{modulation equations} as the following coupled system
\begin{subequations}\label{modulation full}
\begin{align}
&A(t)\dot{\sigma}(t)=F(t),\qquad t>0,\label{modulation ODE}\\
&\partial_tR(\by,t)=JH(t)R(\by,t)+\left(\dot{D}\cdot\partial_\by \right) R(\by,t)+\left(\dot{\xi}+\frac{\dot{\bv}\cdot\by}{2}\right)JR(\by,t)-JN(\by,t)\nonumber\\
&\qquad\qquad+\left(\dot{\xi}+\frac{\dot{\bv}\cdot\by}{2}\right)J\phi_\omega(\by)+\left(\dot{D}\cdot\partial_\by\right)\phi_{\omega}(\by)
-\dot{\omega}\partial_\omega\phi_\omega(\by),\quad \by\in\bR^d,\ t>0,\label{modulation PDE}\\
&\sigma(0)=\left(\omega_0,\ \bv_0,\ D_0,\ \xi_0\right),\quad R(\by,0)=R_0(\by),\quad \by\in\bR^d,
\end{align}
\end{subequations}
with
\begin{align*}
N=\beta\left(|\phi_\omega+R|^2\right)(\phi_\omega+R)-\beta\left(\phi_\omega^2\right)\phi_\omega-\beta\left(\phi_\omega^2\right)R-
2\beta'\left(\phi_\omega^2\right)\phi_\omega^2\left(\begin{matrix}R_1\\
0\end{matrix}\right).
\end{align*}

In the modulation equations, (\ref{modulation ODE}) is a $(2d+2)$-dimensional ODE system and (\ref{modulation PDE}) is a $2$-dimensional dispersive PDE system. They are nonlinear and coupled. The components of $\sigma$ which provide the information of the soliton, i.e. the shape, velocity, frequency and shift, are sometimes referred as collective coordinates in physics \cite{Zhao}. Note that the last component $\xi$ in $\sigma$ does not directly provide the original frequency $\gamma$ of the soliton in (A2). Its derivative $\dot{\xi}$ instead of itself is truly involved in the modulation equations (\ref{modulation full}) and indicates the dynamics of $\gamma$ by (\ref{xi def}), i.e.
 \begin{align}\label{eq:gm}
\dot{\gamma}(t)=\dot{\xi}(t)-\frac{1}{2}\dot{\bv}(t)\cdot\int_0^t\bv(s)\ud s-\frac{1}{2}\dot{\bv}(t)\cdot D(t), \quad t>0.
\end{align}
 Thus to give the complete information of the soliton, besides solving the modulation equations for  $\sigma$, one also needs to solve the ODE (\ref{eq:gm}) at the same time for $\gamma$. In the modulation equations (\ref{modulation full}), everything is real-valued. After solving it, we restore the complex-valued scalar $R=R_1+iR_2$ to give the radiation.

\subsection{Brief review and discussion}
The modulation equations (\ref{modulation full}) in asymptotic orders, have been studied mathematically in \cite{Buslaev,Cuccagna}. It has been pointed out that
when the initial radiation $R_0$ is small, under certain conditions on the nonlinearity $\beta$, the modulations equations (\ref{modulation full}) are well-posed. Moreover, the solution of the modulations equations preserve the orthogonality (\ref{orthog}), and via (A), it gives the solution of the NLS equation (\ref{nls}). When $t\to\infty$, the radiation $R$ will vanish at the same rate as the solution to the linear Schr\"{o}dinger equation with constant coefficients, and the collective coordinates $\sigma$ will turn to reach a steady state.

We remark the modulation equations (\ref{modulation full}) are for describing the multichannel dynamics of a moving soliton and radiation (\ref{multichannel}). For the case of a standing wave and radiation, the modulation equations are different and have been studied in \cite{Soffer1,Zhao} from both theoretical and numerical points of view. For the multiple solitons with mutual interaction and radiation case, the modulation equations and the mathematical analysis will become very complicated, so it has not been studied a lot yet. We refer the readers to \cite{Schlag} for some results.

The method of modulation equations described above is closely related to the method of collective coordinates.
In fact, it was proven in \cite{Soffer1} that if the solution at any time is close to a soliton, then it can be written as a soliton plus small remainder which is also orthogonal in the above sense. Hence, both methods give a decomposition with small corrections.
However, the method of modulation equations also give a PDE for the radiation term, and coupling between the radiation and the ODE's.
Hence the modulation equations approach allows : (i) Control of the error in the ODE's.
(ii) Allows approximating the effect of radiation on the soliton dynamics, either exactly, or by using a good approximation of the coupling term.
(iii) When the radiation effect is critical, as in dissipation mediated processes by the radiation \cite{Soffer2,Soffer3}, one can derive the leading dissipation (radiation mediated!) term from the couples equations, and find the leading behavior for processes in which a soliton changes state, for example( from excited to ground state. see \cite{Soffer3, Soffer4}).
(iv) Resolving the soliton part as it arrives the boundary of the domain of computation. This can not be handled by absorbing boundaries \cite{SStu}.
(v) Allows the rigorous asymptotic stability and scattering over arbitrary large time intervals.

\section{Numerical method}\label{sec: eigen pro}
In this section, we present the numerical methods for solving the modulation equations. To do that, we first write down the numerical algorithm for solving the eigenvalue problem (\ref{eigenfunction}) for the soliton, and then we give the numerical method for the modulation equations (\ref{modulation full}).

\subsection{For the eigenvalue problem}
Since the modulation equations (\ref{modulation full}) replies on the solution of the eigenvalue problem (\ref{eigenfunction}) for all times, so we need to call for a numerical algorithm for firstly finding the nonlinear bound state $\phi_\omega$ from (\ref{eigenfunction})
$$\Delta\phi_\omega(\mathbf{x})+\beta(\phi_\omega^{2}(\mathbf{x}))\phi_\omega(\mathbf{x})=\omega\phi_\omega(\mathbf{x}),\quad \bx\in\bR^d.$$
For a given $\omega>0$, we remark that the above equation is a nonlinear elliptic problem which can be solved numerically by the classical Newton method. However the requirement by the Newton method of an accurate enough initial guess will cause low efficiency in discretizing the modulation equations (\ref{modulation full}) later \cite{Zhao}. This numerical burden becomes severe especially in high dimensions. Besides, more erratic failures of the Newton method for solving the travelling waves have been pointed in \cite{Boyd}. Thus, here we apply either the Petviashvili's iteration method \cite{Petviashvili} or the numerical algorithm which is proposed in \cite{Zhao} in spirit of the generalized Petviashvili's method \cite{Ablowitz,Lakoba}, to solve (\ref{eigenfunction}).

In details, for the pure power nonlinearity case in (\ref{eigenfunction}), i.e. $\beta(\rho)=\lambda\rho^m$ for $\rho\in\bR$ with $\lambda\in\bR,\,m>0$, we apply the standard Petviashvili's iteration method, which reads as the following. Denote $\phi^n_\omega\,(n=0,1,\ldots,)$ as the approximation to $\phi_\omega$ and the Fourier transform
$$\widehat{\phi_\omega^{n}}(\bk)=\int_{\bR^d}\phi_\omega^n(\bx)\fe^{-i\bk\cdot\bx}\ud\bx,
\quad \widehat{f^{n}}(\bk)=\int_{\bR^d}f^n(\bx)\fe^{-i\bk\cdot\bx}\ud\bx,\quad \bk\in\bR^d,$$
where $f^n(\bx):=\lambda(\phi_\omega^{n}(\bx))^{2m+1}$.
Suppose $\phi^0_\omega$ is the initial guess, then for $n\geq0$,
\begin{align}\label{Petmethod S}
\widehat{\phi_\omega^{n+1}}(\bk)=(M^n)^\alpha\frac{\widehat{f^n}(\bk)}{\omega+|\bk|^2},\quad \bk\in\bR^d,
\end{align}
with
\begin{align}\label{Petmethod E}
M^n=\frac{\int_{\bR^d}\left(\omega+|\bk|^2\right)[\widehat{\phi_\omega^{n}}(\bk)]^2\ud \bk}{\int_{\bR^d}\phi_\omega^{n}(\bk)\widehat{f^n}(\bk)\ud\bk},
\end{align}
and $\alpha=\frac{2m+1}{2m}$ which gives the fastest convergence rate of the iteration \cite{Pelinovsky}. The iteration is stopped when
\begin{equation}\label{Cauchy}
|M^n-1|\leq\eps,
\end{equation}
 for some chosen threshold $\eps>0.$

The sequence $\{\phi_\omega^n\}_{n\geq0}$ from the above Petviashvili's method has been proved rigorously to converge to the soliton $\phi_\omega$ in one and two space dimensions in \cite{Pelinovsky}. From the application point of view, the Petviashvili's method has been recognized as the most efficient numerical scheme for computing the solitary waves in a class of problems like (\ref{eigenfunction}) with power nonlinearity and without external potential \cite{Lakoba}.  An improved version is considered in \cite{Demanet}, but it is more involved.

For the practical implementation of the iterative method, since the bound state decays very fast to zero at far field \cite{Cuccagna}, we truncate the problem (\ref{eigenfunction}) onto a finite interval $\Omega=[-L,L]^d$ and impose the periodic boundary condition for numerical issues, i.e.
\begin{subequations}\label{eigen trun}
\begin{align}
&\Delta\phi_\omega(\mathbf{x})+\beta(\phi_\omega^{2}(\mathbf{x}))\phi_\omega(\mathbf{x})=\omega\phi_\omega(\mathbf{x}),\quad
\mathbf{x}\in\Omega,\label{eigen trun a}\\
&\phi_\omega(\mathbf{x}_j)=\phi_\omega(\mathbf{x}_{-j}),\quad
\partial_{x_j}\phi_\omega(\mathbf{x}_j)=\partial_{x_j}\phi_\omega(\mathbf{x}_{-j}),\quad \bx_j,\bx_{-j}\in\partial\Omega,\label{eigen trun b}
\end{align}
\end{subequations}
where $\bx=(x_1,\ldots,x_d)^T$, $\bx_j=(x_1,\ldots,x_{j-1},L,x_{j+1},\ldots,x_d)$ and $\bx_{-j}=(x_1,\ldots, x_{j-1},-L,x_{j+1},\ldots,x_d)$ for $j=1,\ldots,d$. The Fourier transforms and integrations in the Petviashvili's method (\ref{Petmethod S})-(\ref{Petmethod E}) are implemented by means of the Fourier pseudo-spectral method \cite{Shen}. For example, in one dimension $\Omega=[-L,L]$, if we denote $x_j=-L+j\cdot\frac{2L}{N}$ for $j=0,1,\ldots,N$ with $N$ an even integer, then
\begin{equation}\label{fp}
\widehat{\phi_\omega^{n}}(k_l)\approx\frac{2L}{N}\sum_{j=0}^{N-1}\phi_\omega^n(x_j)\fe^{-ik_l(x_j+L)},\quad
\widehat{f^{n}}(k_l)\approx\frac{2L}{N}\sum_{j=0}^{N-1}f^n(x_j)\fe^{-ik_l(x_j+L)},\quad k_l=\frac{\pi l}{L},
\end{equation}
for $l=-N/2,-N/2+1,\ldots,N/2-1,$ and
\begin{equation}\label{integration}
 M^n\approx\frac{\sum_{l=-N/2}^{N/2-1}\left(\omega+k_l^2\right)[\widehat{\phi_\omega^{n}}(k_l)]^2}
{\sum_{l=-N/2}^{N/2-1}\phi_\omega^{n}(k_l)\widehat{f^n}(k_l)}.
\end{equation}
This implementation introduces some extra error. The choice of the $L$ determines the boundary truncation error of the soliton and the error of the quadrature for approximating the integrals as in (\ref{integration}). The choice of the $N$ controls the interpolation error of the Fourier transforms as in (\ref{fp}).
Both error should decay exponentially when $L$ and $N$ increase, if the solution is smooth and decay very fast at the far field.

For general nonlinearity case in (\ref{eigenfunction}), i.e. $\beta(\cdot):\bR\to\bR$ is some general nonlinear function, we apply the numerical algorithm proposed in \cite{Zhao}. By adopting the same notations as above and truncating the problem as (\ref{eigen trun}), the algorithm reads:

\emph{Step 1} Find the ground state of the Hamiltonian functional
\begin{equation*}
H^n(\phi)=\int_\Omega\left[|\nabla\phi|^2-\beta\left((\phi_\omega^n)^2\right)|\phi|^2\right]\ud \mathbf{x},
\end{equation*}
in the unit sphere of $L^2(\Omega)$. Denote the solution as
\begin{equation}\label{algor: eigen1}
\widetilde{\phi}_\omega^{n+1}:= \arg \min\{H^n(\phi):\phi\in L^2(\Omega),\ \|\phi\|_{L^2}=1,\ \phi(x)>0\}.
\end{equation}

\emph{Step 2} Scale the ground state $\widetilde{\phi}_\omega^{n+1}$ according to the energy $\omega$. That is to find the scaling constant $c^n$ such that
\begin{equation}\label{algor: eigen2}
\phi_\omega^{n+1}:=c^n\widetilde{\phi}_\omega^{n+1},
\end{equation}
satisfying
\begin{align*}
\int_\Omega\left[-|\nabla\phi_\omega^{n+1}|^2+\beta\left((\phi_\omega^{n+1})^2\right)(\phi_\omega^{n+1})^2\right]\ud \mathbf{x}=\omega\left\|\phi_\omega^{n+1}\right\|_{L^2}^2,
\end{align*}
which is obtained by taking the inner product of (\ref{eigen trun a}) on both sides with $\phi_\omega$ in $L^2(\Omega)$. Then we can solve the equation
$$\int_\Omega\beta\left(\left(c^n\widetilde{\phi}_\omega^{n+1}\right)^2\right)(\widetilde{\phi}_\omega^{n+1})^2\ud \bx=\omega+\int_\Omega \left|\nabla\widetilde{\phi}_\omega^{n+1}\right|^2\ud \bx,$$
for the value of $c^n$. In particular, when it comes back to the power nonlinearity case $\beta(\rho)=\lambda\rho^{m},\lambda\in\bR,m>0$, we have explicit  formula
\begin{equation*}
c^n=\left|\frac{\omega+\int_\Omega\left|\nabla\widetilde{\psi}_E^{n+1}\right|^2
\ud \mathbf{x}}
{\lambda\int_\Omega\left|\widetilde{\psi}_E^{n+1}\right|^{2m+2}\ud\mathbf{x}}\right|^{\frac{1}{2m}}.
\end{equation*}
Then iterate until $\{\phi^n_\omega\}_{n\geq0}$ converges. The Cauchy criterion is used as the stopping condition, i.e.
\begin{equation}\label{Cauchy1}
\|\phi^{n+1}_\omega-\phi^n_\omega\|_{L^\infty}\leq \eps,
\end{equation}
with some chosen threshold $\eps>0.$

To implement the above algorithm, for the first step, we can use the normalized gradient flow method with a backward Euler Fourier pseudospectral discretization \cite{Bao1,BaoDu} to get the ground state $\widetilde{\phi}_\omega^{n+1}$. For the second step, we use the standard Fourier pseudospectral discretization \cite{Shen} for the spatial derivative and integrations. Once $\phi_\omega$ is obtained, $\partial_\omega\phi_\omega$ can be found out from (\ref{Domega}) with periodic boundary conditions on $\Omega$ by the Fourier pseudospectral discretization again, and so does $\partial_\omega^2\phi_\omega$ in (\ref{d2wphi}).
The rigorous mathematical analysis of the convergence of the above algorithm (\ref{algor: eigen1})-(\ref{algor: eigen2}) has been given in \cite{Zhao}. The algorithm has been shown to be efficient, robust even if the initial guess $\phi_\omega^0$ is quite away from the exact solution, and converge to the $\phi_\omega$ rapidly. We refer the readers to \cite{Zhao} for more details.

\begin{remark}
There are many generalized versions of the Petviashvili's method proposed in the literature in order to deal with the eigenvalue problems involving different kinds of external potentials, the coupled equations case and the general nonlinearity case \cite{Ablowitz,Lakoba,Musslimani}. Most of them are proposed with the ad-hoc approximation. The scheme varies when it comes to different cases and it would need choices of some free parameters. They are shown numerically to converge well for some specific problems, but the general convergence results of them are not available. While, the algorithm (\ref{algor: eigen1})-(\ref{algor: eigen2}) proposed in \cite{Zhao} does not have the ad-hoc regularization. The normalization is done in a precise way each step. The scheme is defined very precisely for general cases and the convergence is guaranteed mathematically in \cite{Zhao}.
This is the reason why we use the algorithm (\ref{algor: eigen1})-(\ref{algor: eigen2}) rather than the others such as the spectral renormalization method \cite{Ablowitz}.
\end{remark}

\begin{remark}
The iteration algorithm (\ref{algor: eigen1})-(\ref{algor: eigen2}) also applies to the pure power nonlinearity case in (\ref{eigen trun}). It appears to more complicated than the Petviashvili's method (\ref{Petmethod S})-(\ref{Petmethod E}), since it is originally derived in \cite{Zhao} with an external potential where the linear differential operator and the potential have to be treated separately in the computation. In that case, it is impossible to stay just in the Fourier frequency space for computing. Another issue that makes (\ref{algor: eigen1})-(\ref{algor: eigen2}) more involved than the (\ref{Petmethod S})-(\ref{Petmethod E}) is the call of an imaginary-time evolution method for obtaining the ground state of linearized Hamiltonian. Although we are using the classical normalized gradient flow method for simplicity, one needs to realize there are many new developed numerical techniques in recent research for rapidly increasing the efficiency of computing the ground state. Thus, a fair comparison between the two iterative methods would need a systematical study, which is beyond the scope of this paper.
\end{remark}

\subsection{For the modulation equations}
Similar to the eigenvalue problem, since the dispersive wave $R$ also decays very fast to zero at far field at finite time, we truncate the problem (\ref{modulation PDE}) onto a finite interval $\Omega=[-L,L]^d$ and impose the periodic boundary condition for computation. The truncated initial boundary value problem of the modulation equations read,
\begin{subequations}\label{modulation trun}
\begin{align}
&A(t)\dot{\sigma}(t)=F(t),\qquad t>0,\label{modulation trun ODE}\\
&\partial_tR(\by,t)=JH(t)R(\by,t)+\left(\dot{D}\cdot\partial_\by \right) R(\by,t)+\left(\dot{\xi}+\frac{\dot{\bv}\cdot\by}{2}\right)JR(\by,t)-JN(\by,t)\nonumber\\
&\qquad\qquad+\left(\dot{\xi}+\frac{\dot{\bv}\cdot\by}{2}\right)J\phi_\omega(\by)+\left(\dot{D}\cdot\partial_\by\right)\phi_{\omega}(\by)
-\dot{\omega}\partial_\omega\phi_\omega(\by),\quad \by\in\Omega,\ t>0,\label{modulation trun PDE}\\
&\sigma(0)=\left(\omega_0,\ \bv_0,\ D_0,\ \xi_0\right),\quad R(\by,0)=R_0(\by),\quad \by\in\Omega,\\
&R(\mathbf{y}_j)=R(\mathbf{y}_{-j}),\quad
\partial_{y_j}R(\mathbf{y}_j)=\partial_{y_j}R(\mathbf{y}_{-j}),\quad \by_j,\by_{-j}\in\partial\Omega,\ j=1,\ldots,d,\label{mudulated: bd}
\end{align}
\end{subequations}
where $\by_j,\by_{-j}$ are defined similar as $\bx_j,\bx_{-j}$.

Choose the time step size $\tau=\Delta t>0$ and denote the time steps by
$t_n:=n\tau, \, n=0,1,\ldots.$ To present the scheme, we denote
\begin{align*}
&\sigma^n=(\omega^n,\bv^n,D^n,\xi^n)\approx \sigma(t_n),\quad  A^n\approx A(t_n),\quad  F^n\approx F(t_n),\\\
& H^n\approx H(t_n),\quad \gamma^n\approx \gamma(t_n),\quad R^n(\mathbf{y})\approx R(\mathbf{y},t_n),\\ &\phi_{\omega}^n(\by)\approx\phi_{\omega(t_n)}(\by),\quad
\partial_\omega\phi_{\omega}^n(\by)\approx\partial_\omega\phi_{\omega(t_n)}(\by),\quad N^n(\by)\approx N(\by,t_n),
\end{align*}
and introduce the finite difference operator on some grid functions $f^n$,
$$\delta_tf^n:=\frac{f^{n+1}-f^{n-1}}{2\tau},\qquad n=1,2,\ldots.$$
Then a semi-implicit leap-frog finite difference temporal discretization of (\ref{modulation trun}) reads,
\begin{subequations}\label{modulation semi-dis}
\begin{align}
&A^n\delta_t\sigma^n=F^n,\quad n=1,2,\ldots,\label{modulation semi-dis: sigma}\\
&\delta_tR^n(\by)=\frac{1}{2}\left(JH^n+\delta_tD^n\cdot\partial_\by\right)\left(R^{n+1}(\by)+R^{n-1}(\by)\right)
+\left(\delta_t\xi^n+\frac{1}{2}\delta_t\bv^n\cdot\by\right)JR^n(\by)\nonumber\\
&\qquad\qquad\ -JN^n(\by)+\left(\delta_t\xi^n+\frac{1}{2}\delta_t\bv^n\cdot\by\right)J\phi_\omega^n(\by)
+\left(\delta_tD^n\cdot\partial_\by\right)\phi_{\omega}^n(\by)\nonumber\\
&\qquad\qquad\ -\delta_t\omega^n\partial_\omega\phi_\omega^n(\by),\qquad \by\in\Omega,\ \ n=1,2,\ldots,\label{modulation semi-dis: R}
\end{align}
\end{subequations}
with initial values
$$\sigma^0=\left(\omega_0,\ \bv_0,\ D_0,\ \xi_0\right),\quad R^0(\by)=R_0(\by).$$
Also, together with (\ref{modulation semi-dis}) we dicretize (\ref{eq:gm}) as
$$\delta_t\gamma^n=\delta_t{\xi}^n-\frac{\tau}{4}\sum_{m=0}^{n}\delta_t{\bv}^n\cdot\bv^m-\frac{1}{2}\delta_t{\bv}^n\cdot D^n,
\quad n=0,1,\ldots,$$
where we apply the
composite trapezoidal rule to approximate the integral term.
Since (\ref{modulation semi-dis}) is a two-level scheme, we also need the starting values at $t=t_1$. In order to get a second order accuracy in temporal  approximations, they are obtained by the Taylor's expansion of the solution and noticing the equations (\ref{modulation semi-dis: R}) as
\begin{align*}
&\sigma^1=\sigma^0+\tau\delta_t\sigma^0,\qquad\delta_t\sigma^0=(A^0)^{-1}F^0,\\
&\gamma^1=\gamma^0+\tau\delta_t\gamma^0,\qquad R^1(\by)=R^0(\by)+\tau\delta_tR^0(\by),\\
&\delta_tR^0(\by)=\left(JH^0+\delta_tD^0\cdot\partial_\by\right)R^{0}(\by)
+\left(\delta_t\xi^0+\frac{1}{2}\delta_t\bv^0\cdot\by\right)JR^0(\by)\nonumber\\
&\qquad\qquad\ -JN^0(\by)+\left(\delta_t\xi^0+\frac{1}{2}\delta_t\bv^0\cdot\by\right)J\phi_\omega^0(\by)
+\left(\delta_tD^0\cdot\partial_\by\right)\phi_{\omega}^0(\by) -\delta_t\omega^0\partial_\omega\phi_\omega^0(\by).
\end{align*}
(\ref{modulation semi-dis}) is the semi-discretization of (\ref{modulation trun}). To get the full discretization, i.e. to discretize the space and approximate the above spatial derivatives, we use the standard Fourier pseudospectral method \cite{Shen}. Thus, our numerical method can be referred as the semi-implicit Fourier pseudospectral (SIFP) method.
Here, $\phi_\omega^n$ is obtained by algorithm (\ref{algor: eigen1})-(\ref{algor: eigen2}) from (\ref{eigen trun}) and $\partial_\omega\phi_\omega^n$ is given by (\ref{Domega}).

The SIFP method is clearly time symmetric. In the scheme of SIFP, (\ref{modulation semi-dis: sigma}) is fully explicit, while (\ref{modulation semi-dis: R}) is semi-implicit. So at each time level $t=t_n$, we apply a linear solver, for example the Gauss-Seidel method \cite{Golub}, to get $R^{n+1}$.  We remark that here the reason why we put a time average on the function behind the operator $JH+\dot{D}\cdot\partial_\by$ in (\ref{modulation semi-dis: R}) is to get rid of the stability problems \cite{Bao1}.
Finally, we would like to comment that this modulation equations approach as a numerical solver for the NLS equation works in any space dimensions, since the modulation equations are consistent with the NLS and the corresponding numerical discratization are well-defined in any dimensions.

\begin{remark}
We remark that in the algorithm (\ref{algor: eigen1})-(\ref{algor: eigen2}) for the eigenvalue problem (\ref{eigen trun}) in Section \ref{sec: eigen pro} and the temporally discretized modulation equations (\ref{modulation semi-dis}), one can also use the finite difference method for spatial discretizations. Here we choose the Fourier pseudospectral method for a high accuracy purpose in the case of periodic boundary conditions (\ref{eigen trun b}) and (\ref{mudulated: bd}).
\end{remark}

\begin{remark}
In \cite{Zhao}, the imposed boundary condition is the zero boundary where corresponding sine spectral method is applied, while here we use the periodic boundary condition (\ref{eigen trun b}) and (\ref{mudulated: bd}). Both types of boundary conditions are fine to use as approximations to the physical model after domain truncation if the finite domain is chosen large enough. The reason why we use the periodic boundary condition here is because we still want to apply the spectral method in the presence of the gradient operator in (\ref{modulation trun PDE}).
\end{remark}

\section{Numerical results}\label{sec: result}
In this section, we present the numerical results of modulation equations via using the proposed numerical method. To do that, we first test the correctness and accuracy of the SIFP method. Then we apply the SIFP method to numerically explore the multichannel dynamics.

\subsection{Accuracy test}\label{subsec accuracy}

For simplicity, we consider the one-dimensional case, i.e. $d=1$ and
$$\bx=x,\quad \bv=v,\quad \by=y,$$
in the NLS equation (\ref{nls}) and the modulation equations (\ref{modulation full}), to test the SIFP method (\ref{modulation semi-dis}). We take the cubic nonlinearity, i.e.
\begin{equation}\label{cubic}
\beta(|u|^2)u=\lambda|u|^2u,\qquad \lambda=2,
\end{equation}
in (\ref{nls}) and choose the computation domain $\Omega=[-L,L]$ for the variable $y$ in (\ref{modulation trun}) with $L=16$ which is large enough to ignore the boundary truncation error before the wave reaches the boundary of the $y$-domain during a short time computing.
We choose the initial data in (\ref{modulation: ini})
$$v_0=0.5,\quad \omega_0=1,\quad \gamma_0=1,\quad D_0=0,$$
where we know explicitly the corresponding $\phi_{\omega_0}=\sech(x)$. Then in order to satisfy the orthogonality conditions (\ref{orthog}), we choose
$R(x,0)=R_0(x)=R_1(x,0)+iR_2(x,0)$ as
\begin{align}\label{R ini}
R_1(x,0)=0.2x\fe^{-x^2}-\frac{<0.2 x\fe^{-x^2},x\phi_{\omega_0}>}{\|x\phi_{\omega_0}\|_{L^2}^2} x\phi_{\omega_0},\quad R_2(x,0)=0.
\end{align}
The threshold $\eps$ used for the stopping criterion (\ref{Cauchy}) for the iteration algorithm (\ref{algor: eigen1})-(\ref{algor: eigen2}) is chosen as $\eps=10^{-9}$.

To show the modulation equations with SIFP solve the NLS equation (\ref{nls}) correctly, we solve the modulation equations (\ref{modulation trun}) numerically by the SIFP (\ref{modulation semi-dis}) to get $\phi_{\omega^M}(y),$ $R^M(y)$, $v^M$, $\gamma^M$, $D^M$ and $\omega^n$ for $0\leq n\leq M=T/\tau$, and use the ansatz (A) to construct the numerical solution $u^M(x)$ of the NLS equation (\ref{nls}) with application of the composite trapezoidal rule to approximate the integrals, i.e.
\begin{align*}
u^M(\mathbf{x}):=&\exp\left(i\left(\frac{v^Mx}{2}-\sum_{n=0}^M\frac{\tau|v^n|^2}{8}+\sum_{n=0}^M\frac{\tau\omega^n}{2}
+\gamma^M\right)\right)\nonumber\\
&\times\left[\phi_{\omega^M}\left(x-\sum_{n=0}^M\frac{\tau v^n}{2}-D^M\right)+R^M\left(x-\sum_{n=0}^M\frac{\tau v^n}{2}-D^M\right)\right].
\end{align*}
Then, we compute the error
\begin{equation}\label{e_Phi}
e_{u}(x,T):=u(x,T)-\Phi^M(x),\quad x\in\Omega,
\end{equation}
where the exact solution $u(x,T)$ of the NLS equation (\ref{nls}) is obtained by classical numerical methods such as the time-splitting Fourier spectral method \cite{Bao1,BaoMarkowich} with very small step size, e.g. $\tau=10^{-4},\ h=1/16$.
We test the temporal and spatial discretization errors of the SIFP method separately. Firstly, for the discretization error in time, we take a fine mesh size $h=1/8$ such that the error from the discretization in space is negligible compared to the temporal discretization error. The errors (\ref{e_Phi}) under maximum norm are presented at $T=1$ in Tab. \ref{tab: error temp}. Secondly, for the discretization error in space, we take a very small time step $\tau=10^{-4}$  such that the error from the discretization in time is negligible compared to the spatial discretization error. The corresponding errors under maximum norm are presented at $T=1$ as well and tabulated in Tab. \ref{tab: error spat}.

\begin{table}[h!]
  \caption{The temporal error and convergence rate of the SIFP method for the modulation equation under different time step $\tau$ with $h=1/8$ at $T=1$.}\label{tab: error temp}
  \vspace*{-10pt}
\begin{center}
\def\temptablewidth{1\textwidth}
{\rule{\temptablewidth}{0.75pt}}
\begin{tabular*}{\temptablewidth}{@{\extracolsep{\fill}}llllllll}
                                 & $\tau_0=0.1$       &  $\tau_0/2$          & $\tau_0/2^2$           & $\tau_0/2^3$ &$\tau_0/2^4$ &$\tau_0/2^5$\\[0.25em]
\hline
$\left\|e_{u}\right\|_{L^\infty}$  & 8.70E-3	&3.20E-3	       &1.20E-3	        &3.78E-4       &9.76E-5 &2.41E-5\\
rate                              & --	        & 1.44	           &1.46	        &1.66 &1.96 &2.01\\
\end{tabular*}
{\rule{\temptablewidth}{0.75pt}}
\end{center}
\end{table}

\begin{table}[h!]
  \caption{The spatial error of the SIFP method for the modulation equation under different mesh size $h$ with $\tau=10^{-5}$ at $T=1$.}\label{tab: error spat}
  \vspace*{-10pt}
\begin{center}
\def\temptablewidth{1\textwidth}
{\rule{\temptablewidth}{0.75pt}}
\begin{tabular*}{\temptablewidth}{@{\extracolsep{\fill}}llllll}
                                 & $h_0=1$       &  $h_0/2$          & $h_0/2^2$           & $h_0/2^3$ \\[0.25em]
\hline
$\left\|e_{u}\right\|_{L^\infty}$  & 2.40E-2	&8.03E-5	       &4.02E-8	        &2.55E-9      \\
\end{tabular*}
{\rule{\temptablewidth}{0.75pt}}
\end{center}
\end{table}

Clearly, from Tabs. \ref{tab: error temp}-\ref{tab:   error spat}, we can conclude that the SIFP method (\ref{modulation semi-dis}) solves the multichannel solutions based on the modulation equations for the NLS equation (\ref{nls}) correctly and accurately. The numerical
method has second order accuracy in time and spectral accuracy in space.

\subsection{Comparisons}
\begin{figure}[h!]
{\centerline{\psfig{figure=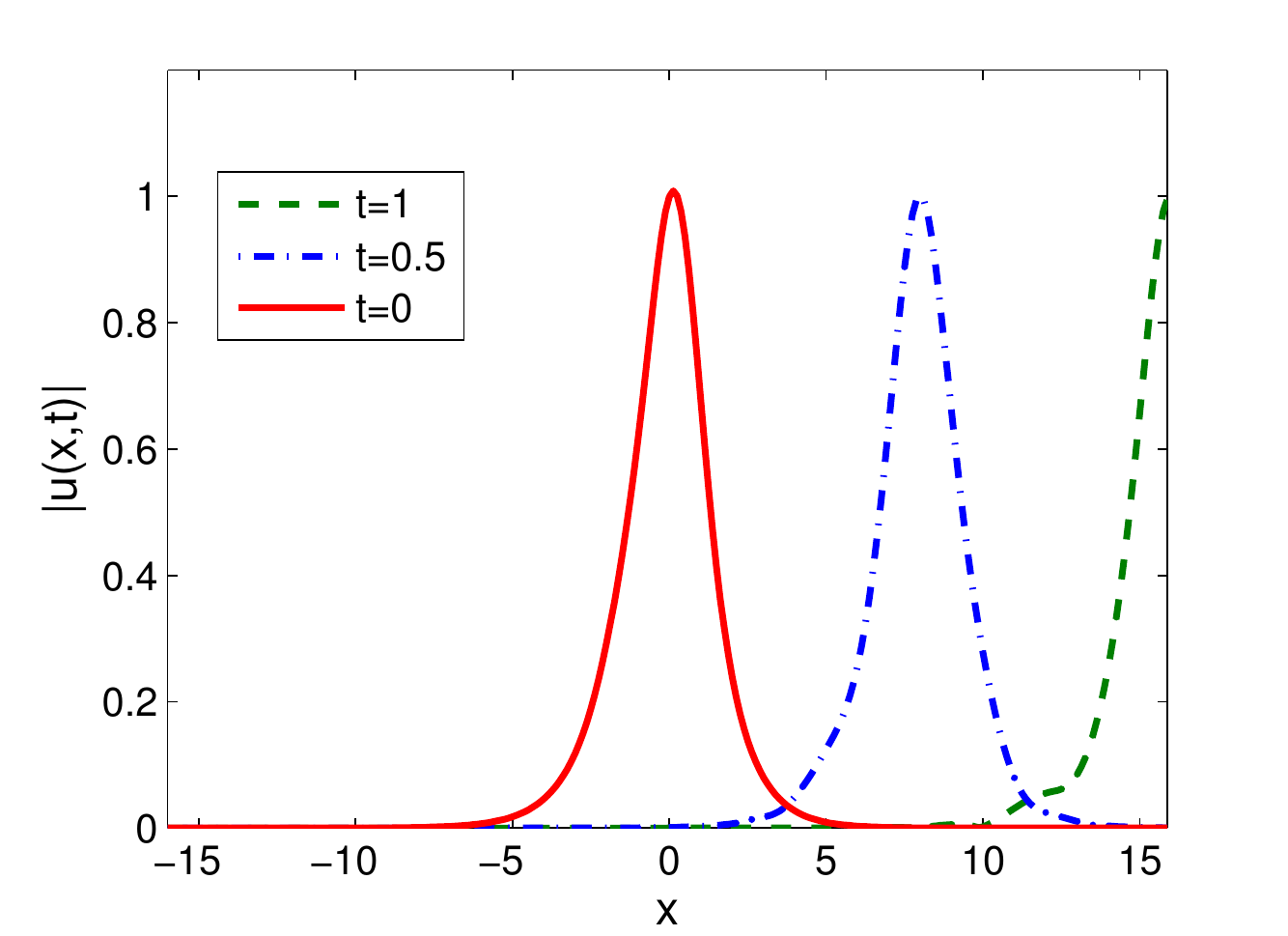,height=4.5cm,width=11cm}}
\
$$
\begin{array}{cc}
\psfig{figure=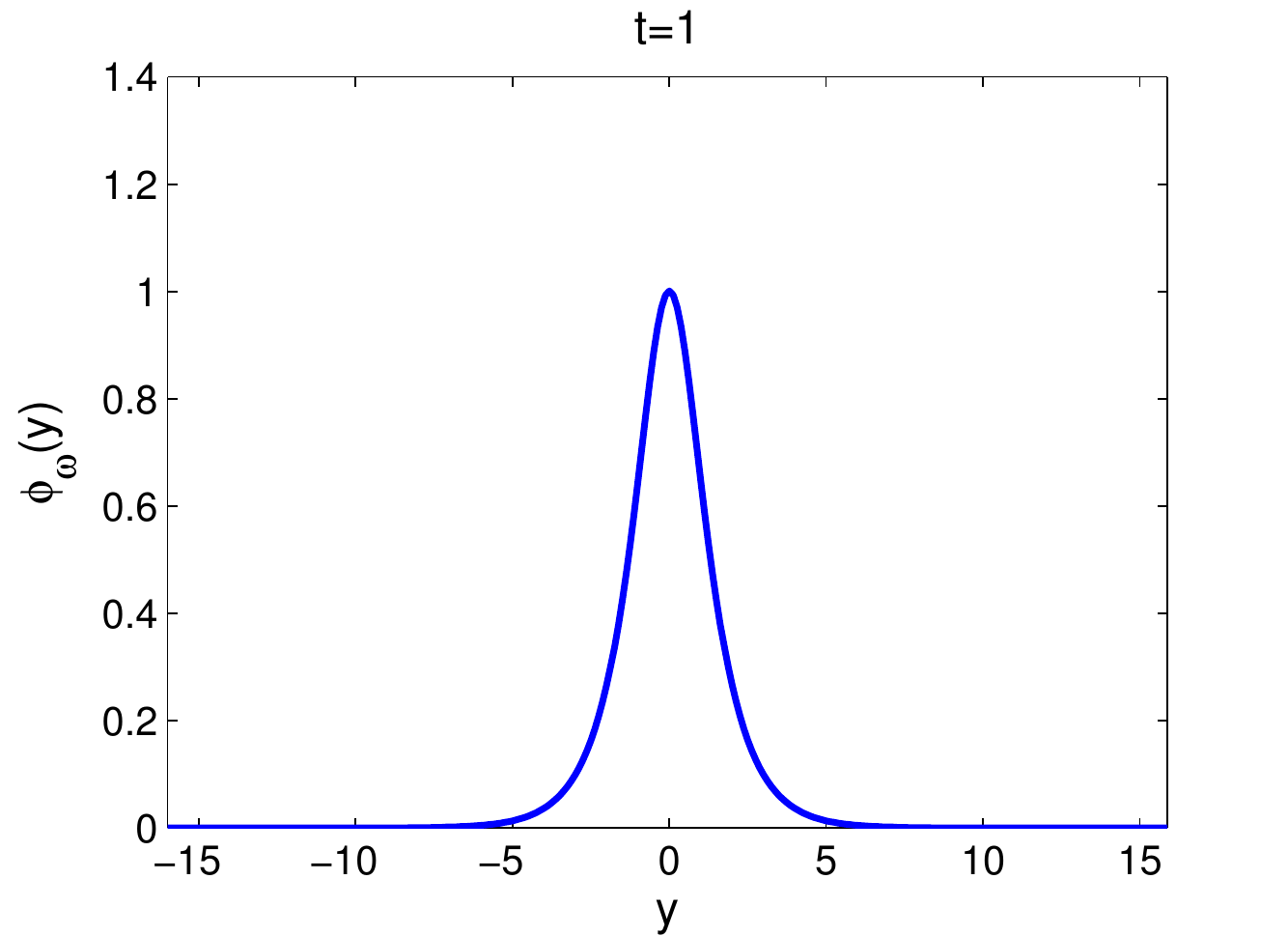,height=5.5cm,width=5.7cm}&\psfig{figure=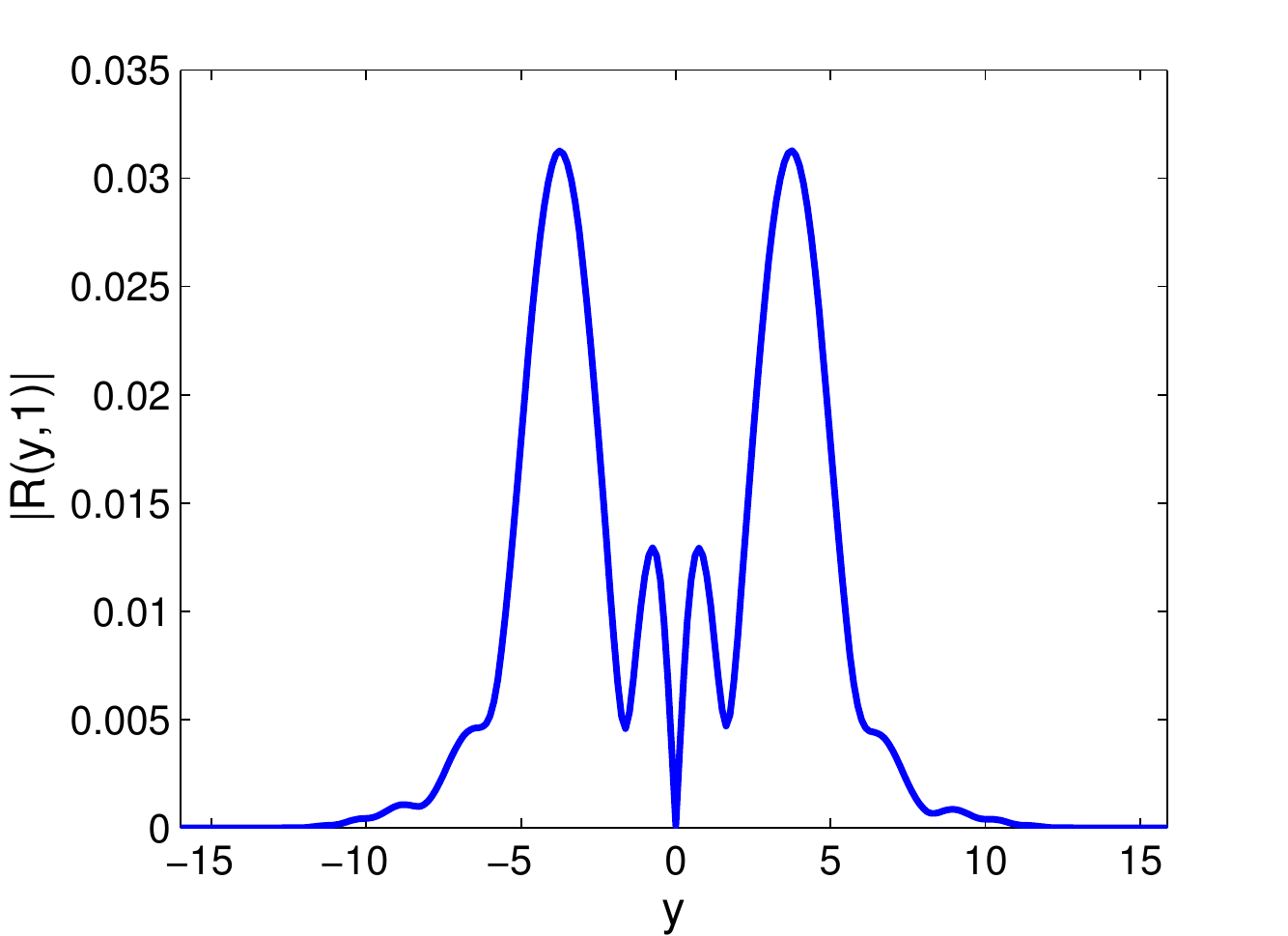,height=5.5cm,width=5.7cm}
\end{array}
$$
}
\caption{Dynamics of the solution $|u(x,t)|$ to the NLS equation in $x$-domain and the $\phi_\omega(y)$ and $R(y,t)$ at $t=1$ in the $y$-domain. }\label{fig:ME}
\end{figure}

\begin{figure}[h!]
{\centerline{\psfig{figure=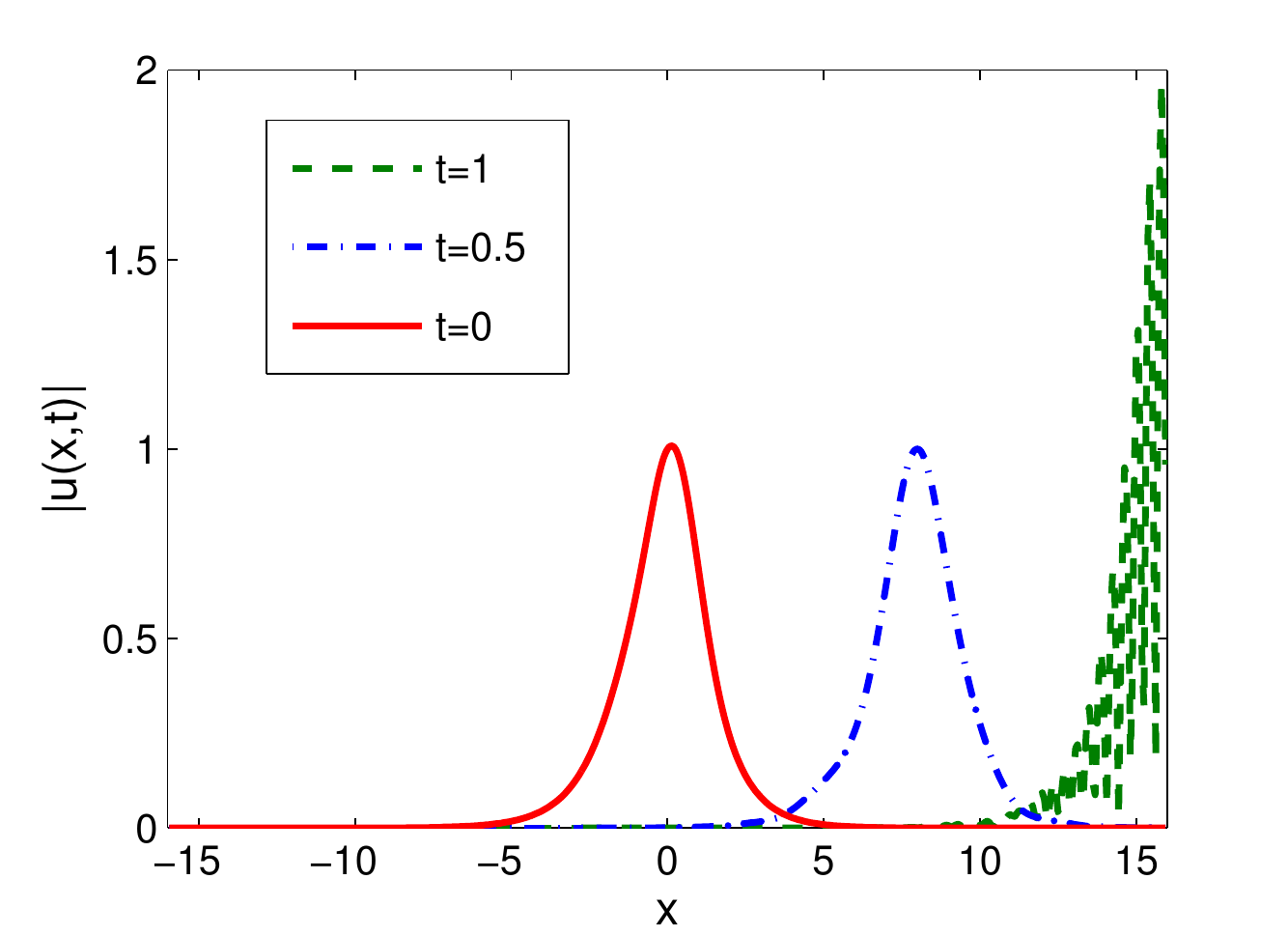,height=4.5cm,width=11cm}}
}
\caption{The solution $|u(x,t)|$ to the NLS equation in the $x$-domain $\Omega=[-16,16]$ by using zero boundary condition and the time-splitting spectral method. }\label{fig:TSSP}
\end{figure}

Now we compare the modulation equations approach to solve the dynamics in the NLS equation with existing direct numerical methods toward discretizing (\ref{nls}). We use the same setup as in (\ref{subsec accuracy}) but with a larger velocity $v_0=16.$ We work on the fixed computational domain $\Omega=[-16,16]$ by using the SIFP method for the modulation equations (\ref{modulation trun}) with the translated variable $y$. The dynamics of the multichannel solution in the original $x$-domain are shown in Fig. \ref{fig:ME}. In Fig. \ref{fig:TSSP}, the same NLS equation (\ref{nls}) problem is solved directly by imposing zero boundary condition and using the time-splitting sine spectral method \cite{Bao1,BaoMarkowich}.

Based on Figs. \ref{fig:ME} and \ref{fig:TSSP}, we can see that on a domain of the same size, within the computational time, the radiation wave in the modulation equations has not reached the boundary yet in the $y$-domain, but the solution of the NLS equation has already hit the boundary due to the velocity.  The waves in the modulation equations approach can pass through the boundary of the $x$-domain naturally, while the waves in the NLS equation are destroyed by the zero boundary condition and the direct PDE solver on $x$-domain. We remark that even if the radiation wave reaches the boundary of $y$-domain, we can use the absorbing boundary techniques to the equation (\ref{modulation trun PDE}) of $R$ and improve the results.

Of course for this simple case, one can change to the periodic boundary condition to avoid the breakdown of the wave at the boundary. The following example illustrate the fails of the periodic boundary condition.

\begin{figure}[h!]
{
$$
\begin{array}{cc}
\psfig{figure=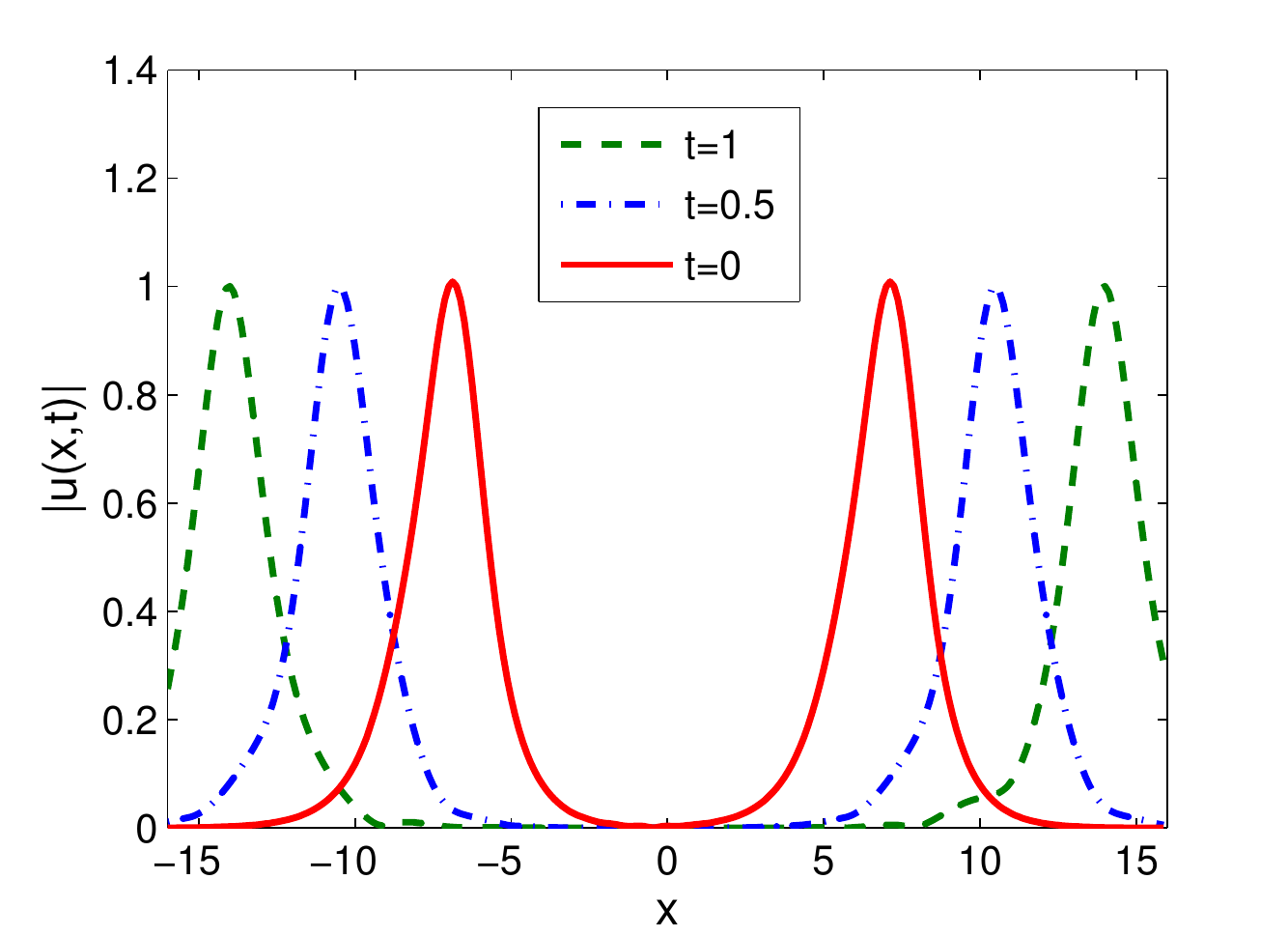,height=5.5cm,width=6.2cm}&\psfig{figure=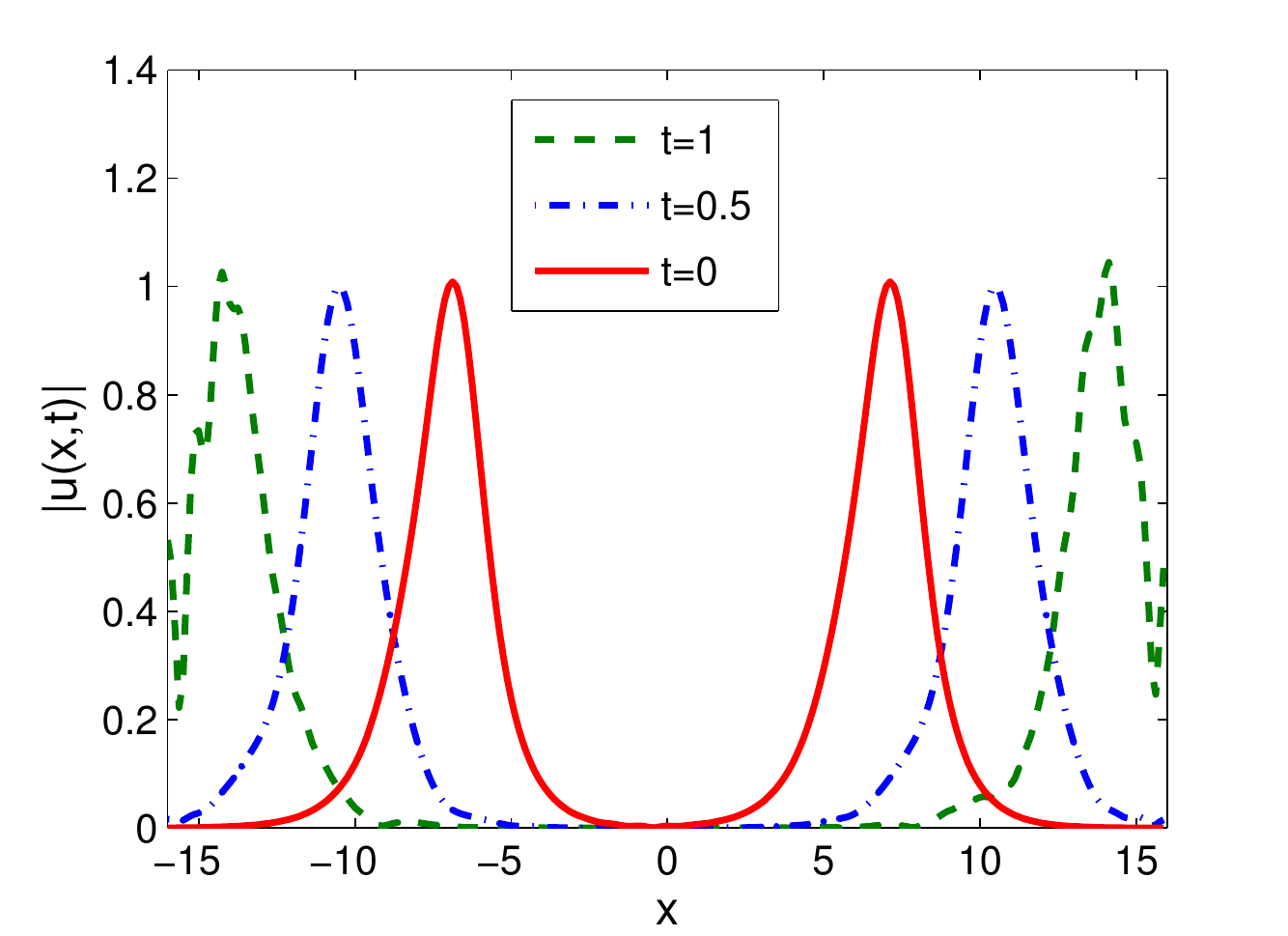,height=5.5cm,width=6.2cm}
\end{array}
$$
}
\caption{Dynamics of the solution $|u(x,t)|$ to the NLS equation in $x$-domain: The results of the modulation equations method (left figure); The results of time-splitting spectral method (right figure). }\label{fig:2soliton}
\end{figure}

When there are two solitons in the multichannel dynamics and two solitons are well-separate initially and moving towards opposite directions, the interactions between the solitons can be ignored. That is to say they can be treated as two single solitons in the NLS equation which can be handled by the modulation equations (\ref{modulation full}). Here we choose the same numerical example but with the velocity and shift
$$v_0=\pm7,\qquad D_0=\pm 7,$$
then the dynamics of the multichannel solution is shown in Fig. \ref{fig:2soliton}. We also show the corresponding results of solving the NLS equation directly by imposing periodic boundary condition and using the time-splitting Fourier spectral method.

By this numerical experiment, based on the results in Fig. \ref{fig:2soliton}, we can see that the modulation equations approach can simulate the multiple solitons case if the solitons are always well separated during the dynamics. However the direct solver for the NLS equation with periodic boundary condition can not, because the waves will collapse when they hit the boundary and enter the domain from the other side.

Now we apply the SIFP method to study the dynamics of the multichannel solutions with the same setup as in (\ref{subsec accuracy}) numerically. In order to provide a long time simulation, we enlarge the domain to $\Omega=[-64,64]$, such that the dispersive wave $R$ is always away from the boundary during the simulation. Take step size $\tau=1E-3,\ h=1/8$, we solve the modulation equation (\ref{modulation trun}) till the collective coordinates $\sigma(t)$ reach the steady state. The dynamics of the collective coordinates $v(t)$, $D(t)$, $\omega(t)$ and $\gamma(t)$ are shown in Fig. \ref{fig2}. The profiles of the dispersive wave $R(y,t)$ at different time are shown in Fig. \ref{fig3}.
\subsection{Explorations on multichannel dynamics}
\begin{figure}[t!]
{$$
\begin{array}{cc}
\psfig{figure=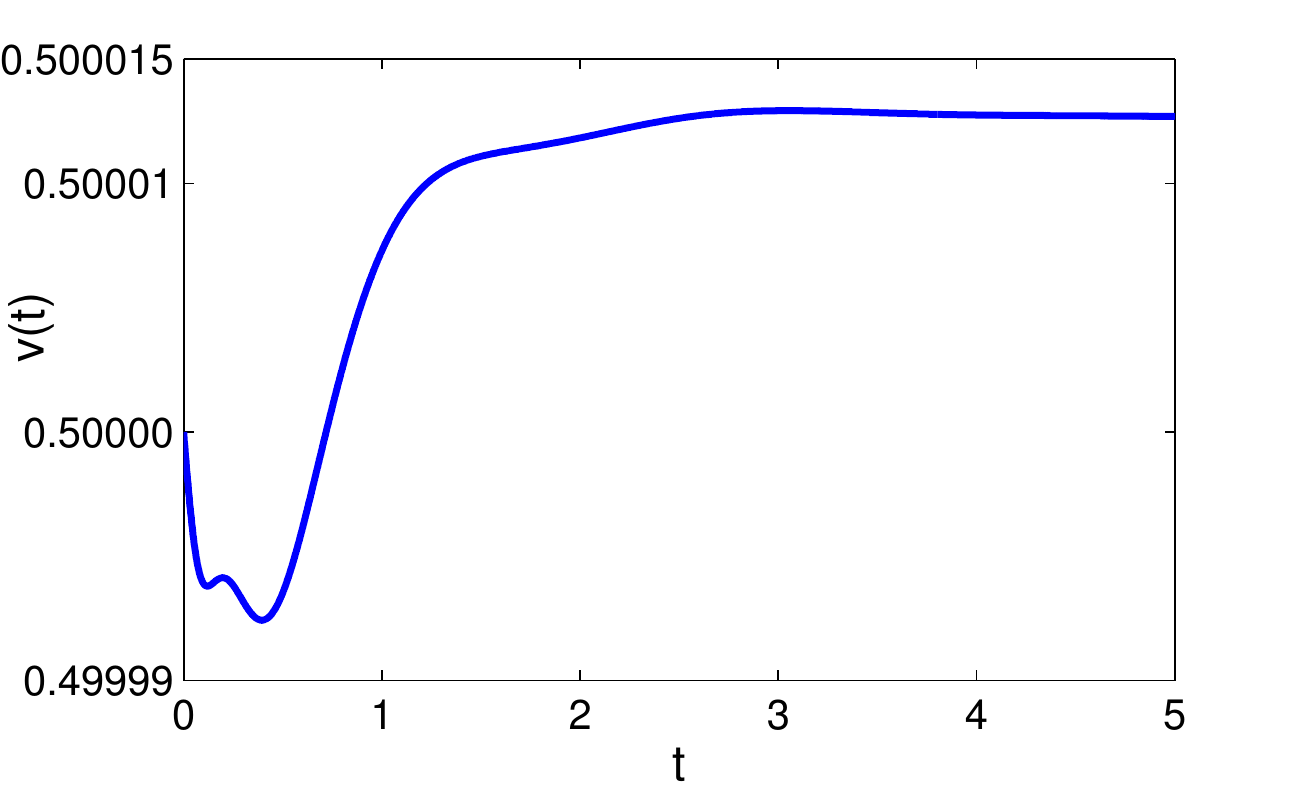,height=5cm,width=6cm}&\psfig{figure=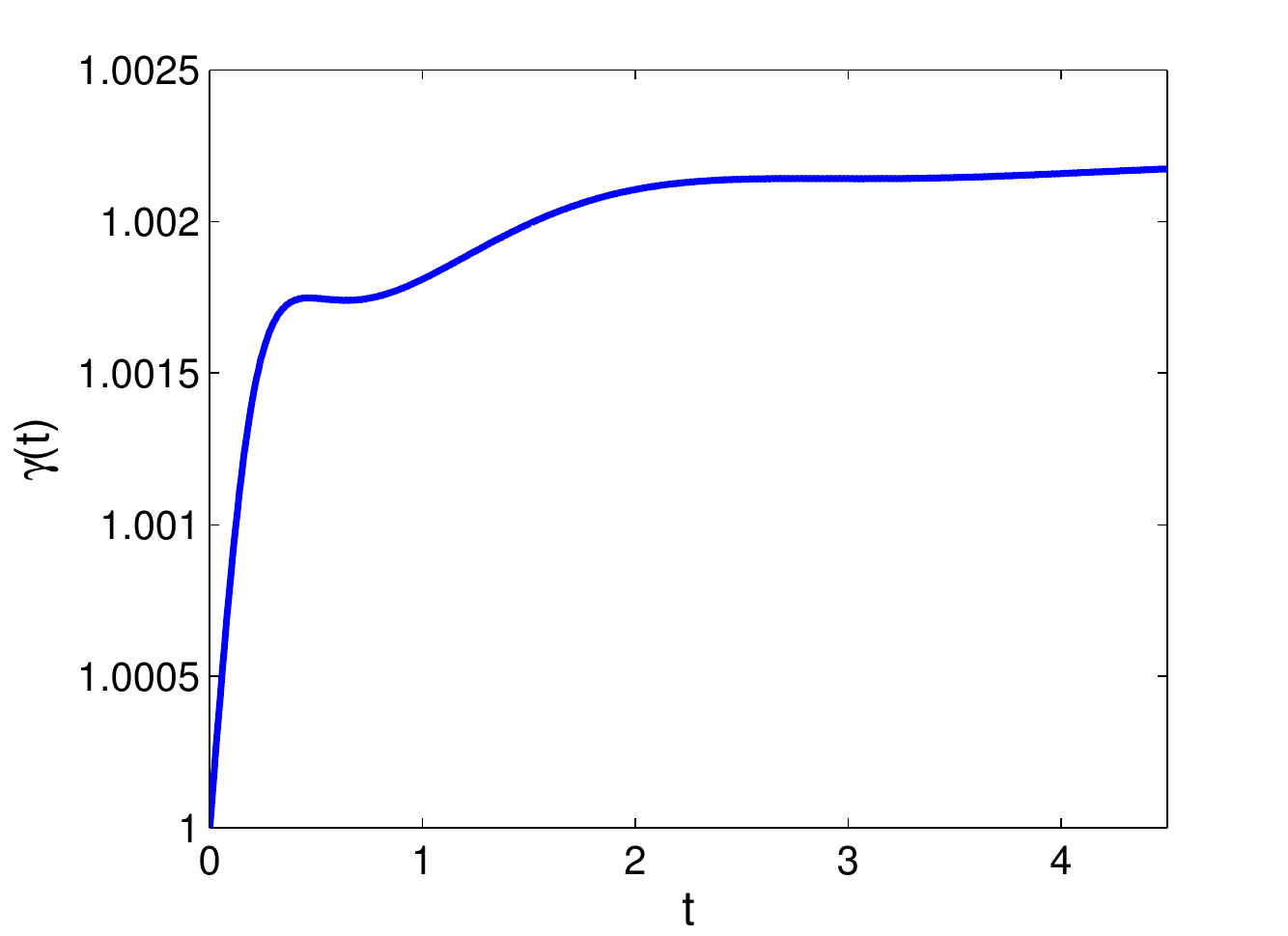,height=5cm,width=6cm}\\
\psfig{figure=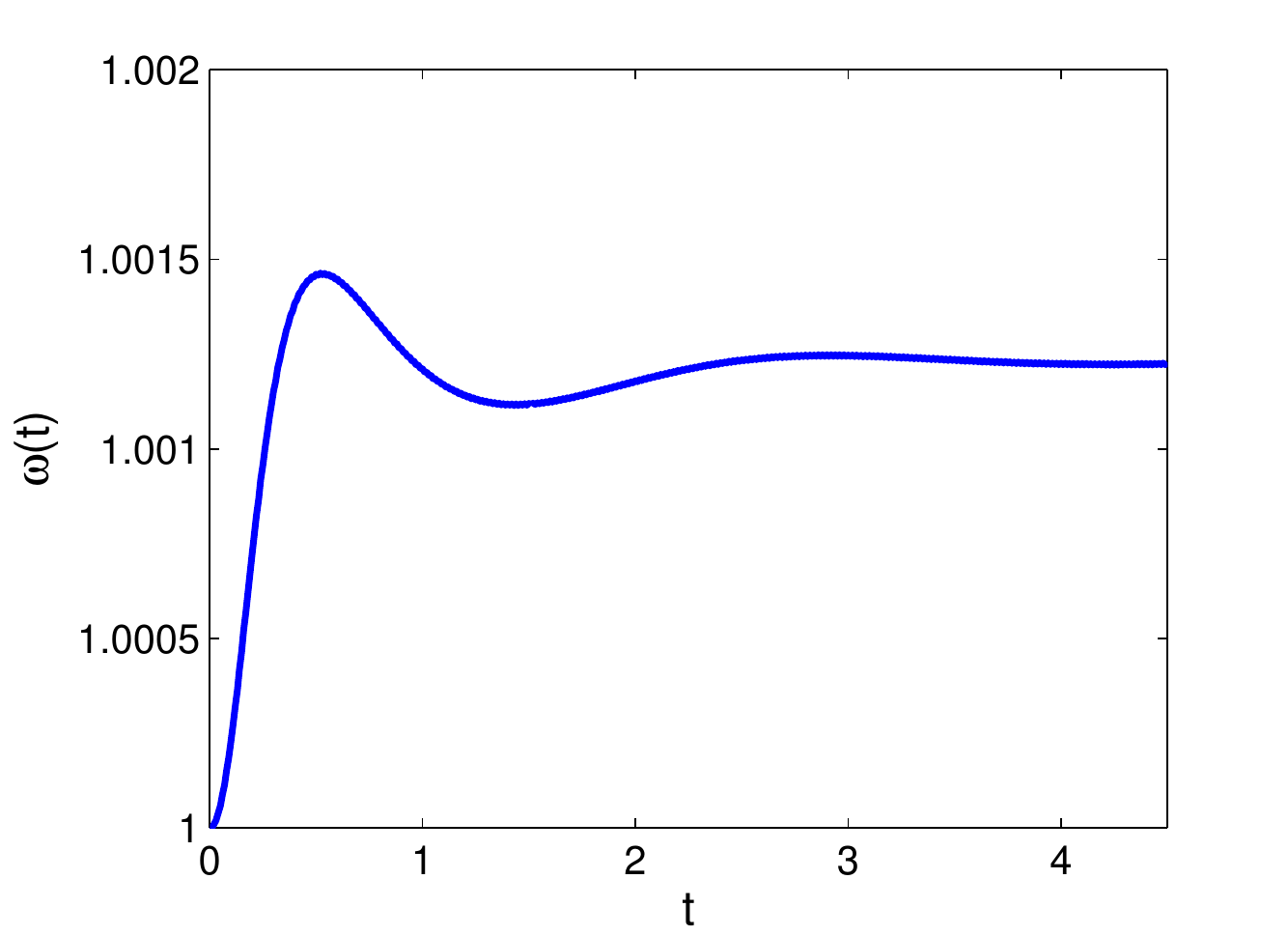,height=5cm,width=6cm}&\psfig{figure=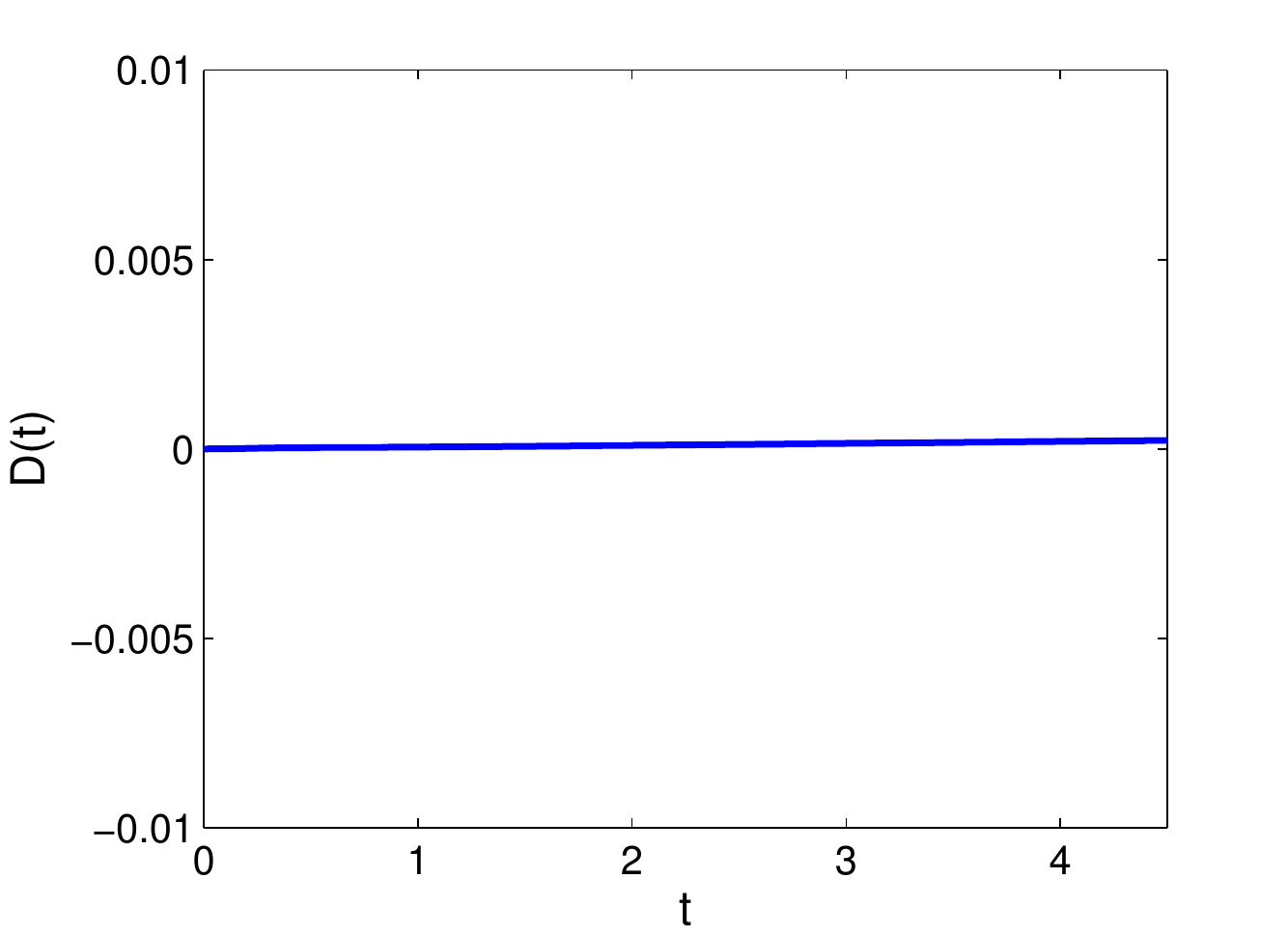,height=5cm,width=6cm}
\end{array}
$$
}
\caption{Dynamics of the collective coordinates $\sigma(t)$. }\label{fig2}
\end{figure}

\begin{figure}[h!]
{\centerline{\psfig{figure=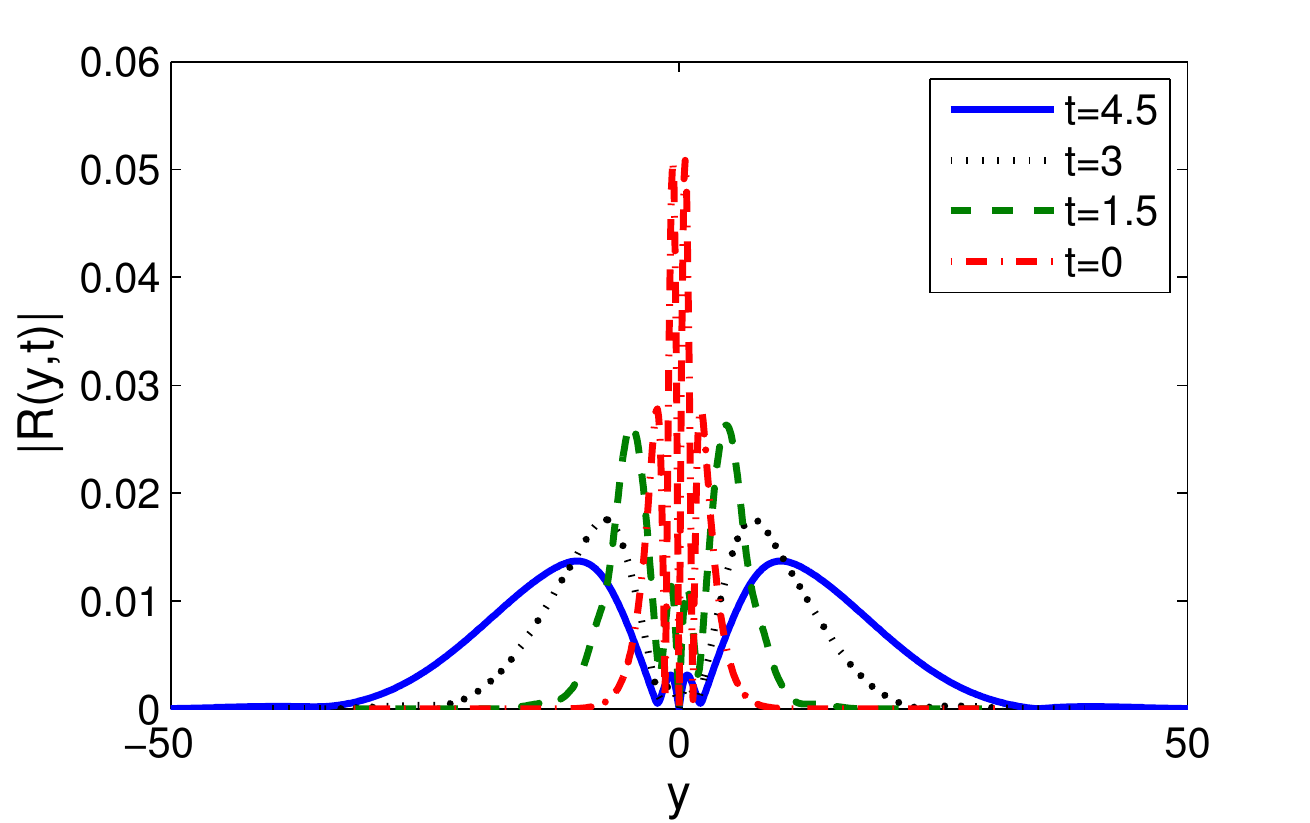,height=5.0cm,width=12cm}}}
\caption{Profiles of the dispersive wave $|R(y,t)|$ at different $t$. }\label{fig3}
\end{figure}

\begin{figure}[t!]
{$$
\begin{array}{cc}
\psfig{figure=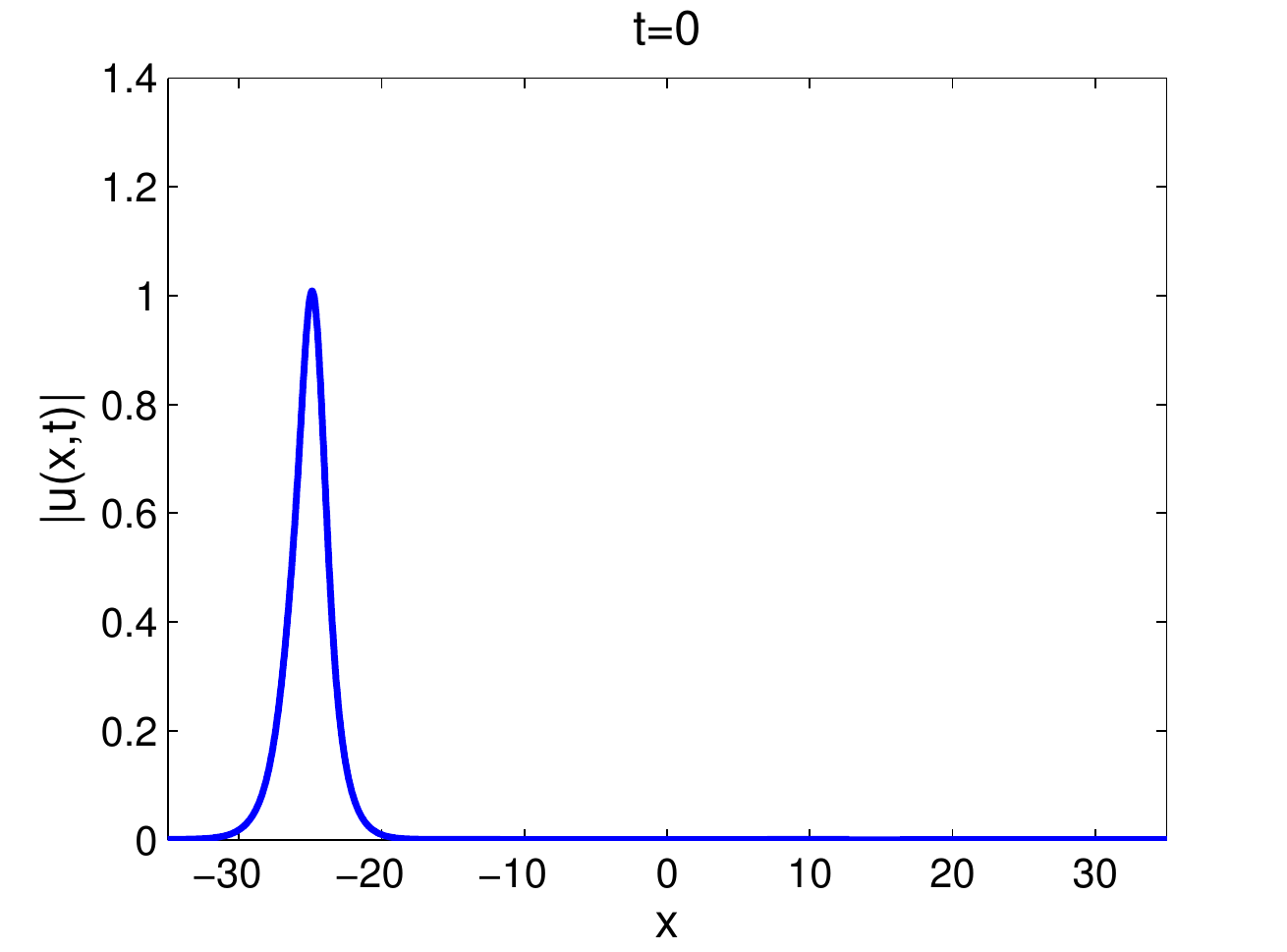,height=3.5cm,width=6cm}&\psfig{figure=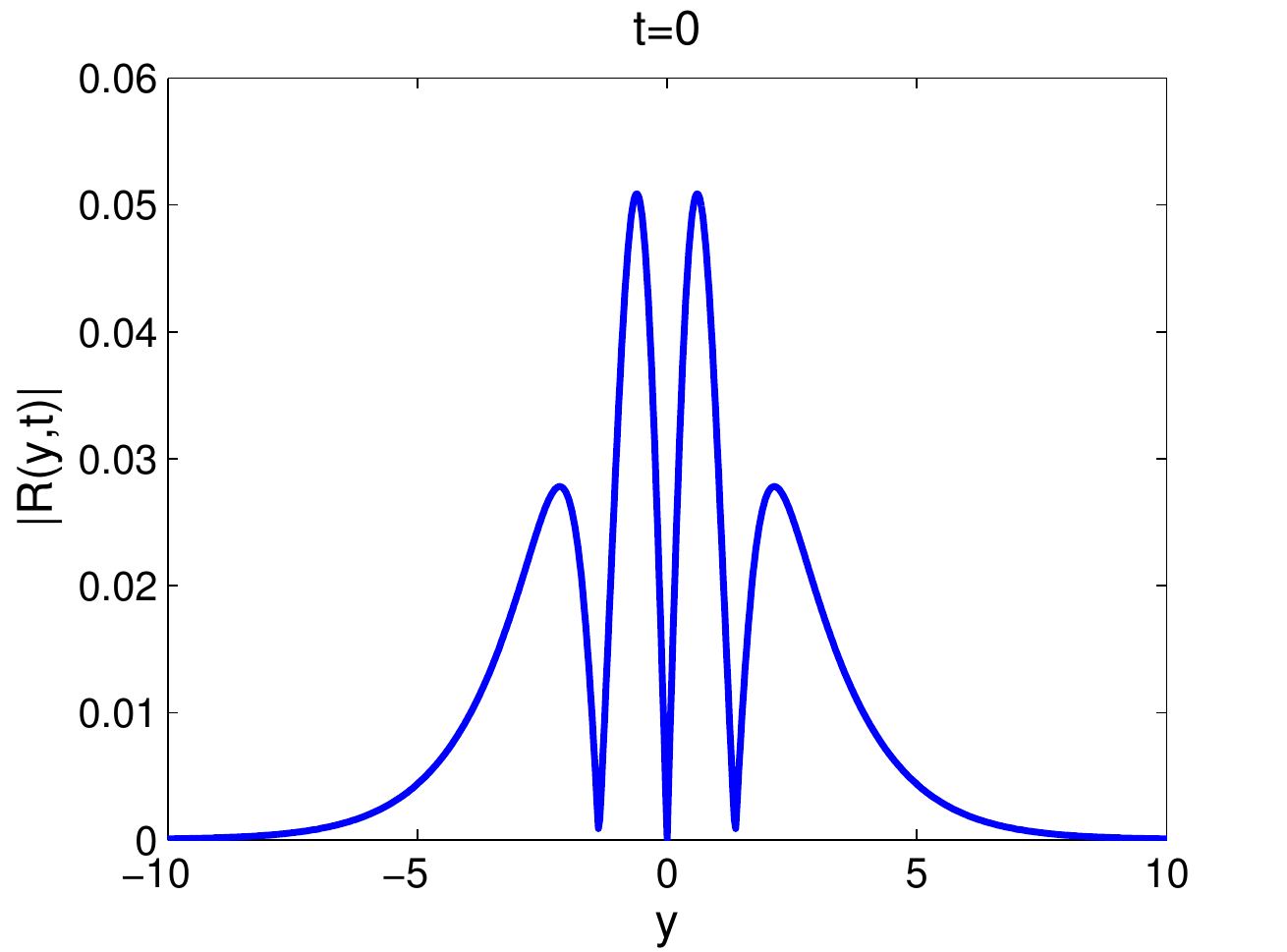,height=3.5cm,width=6cm}\\
\psfig{figure=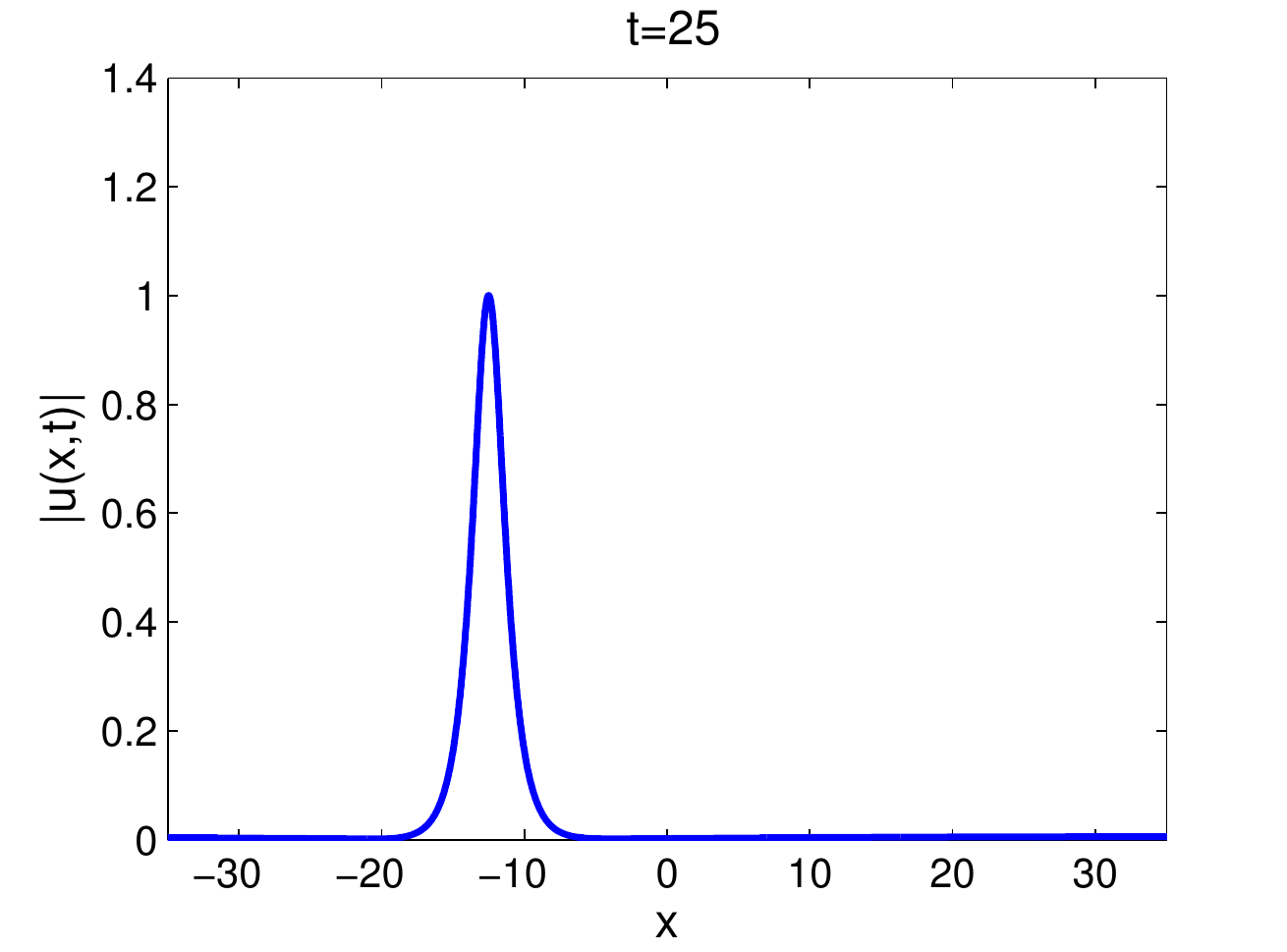,height=3.5cm,width=6cm}&\psfig{figure=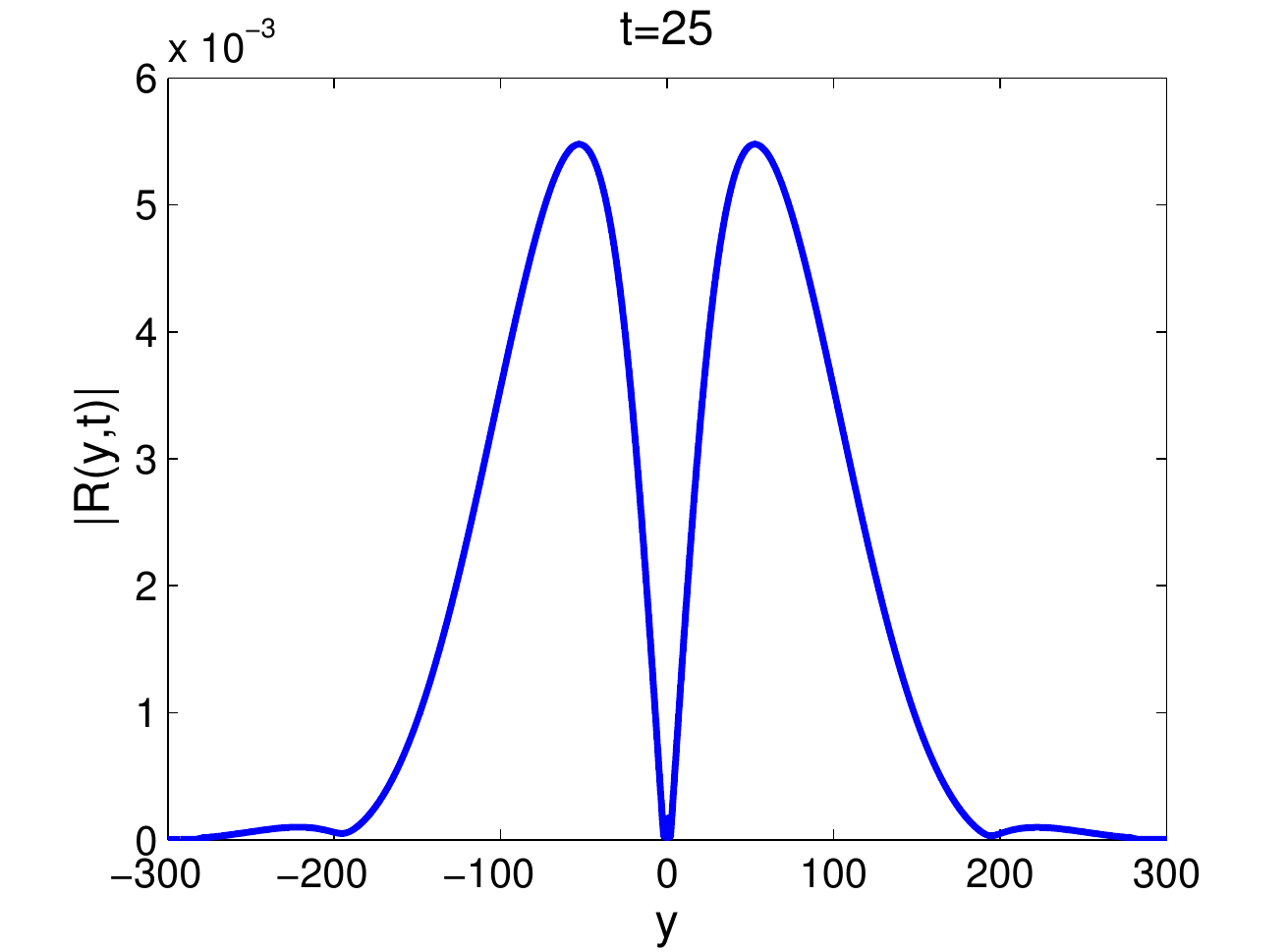,height=3.5cm,width=6cm}\\
\psfig{figure=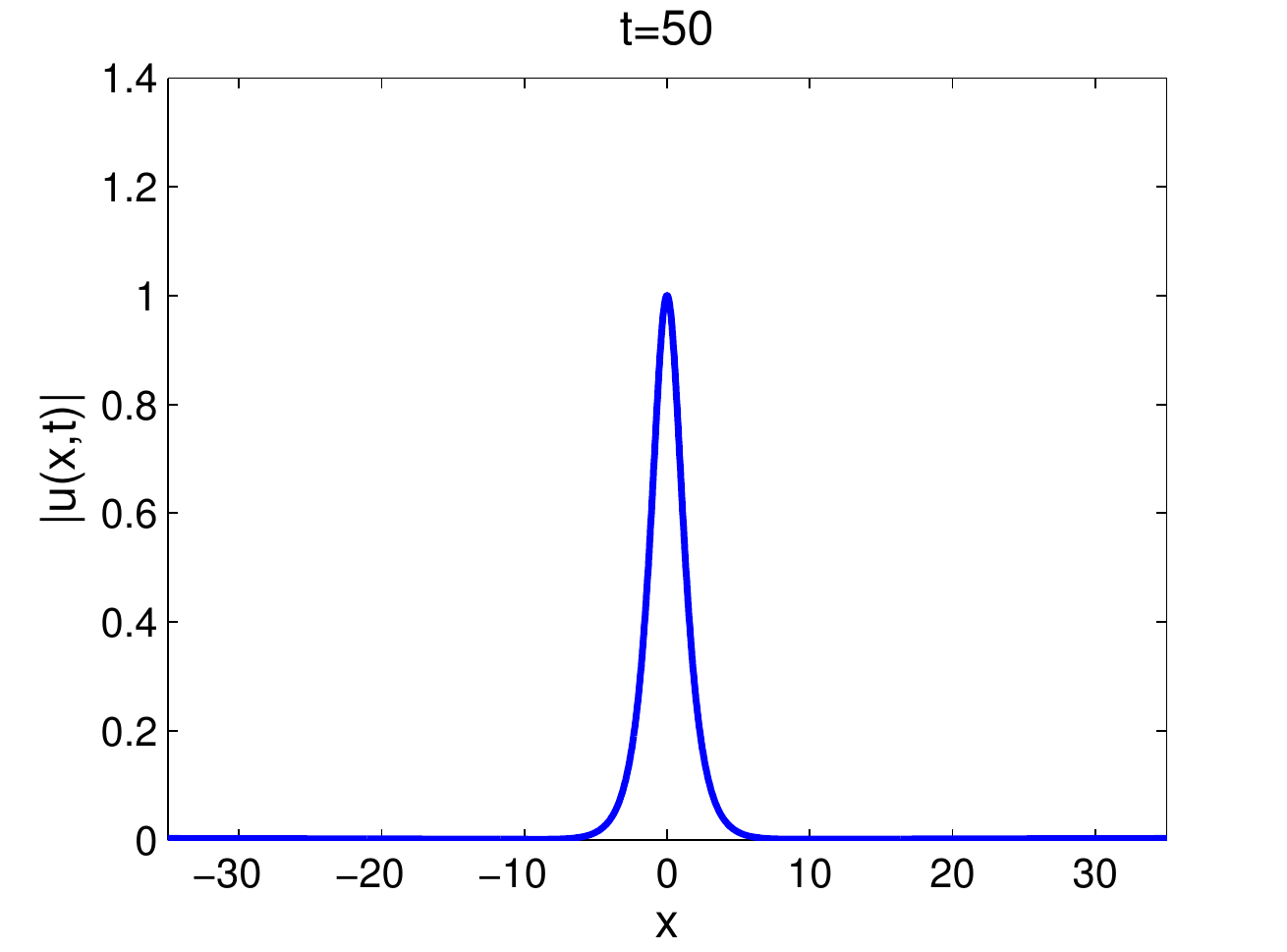,height=3.5cm,width=6cm}&\psfig{figure=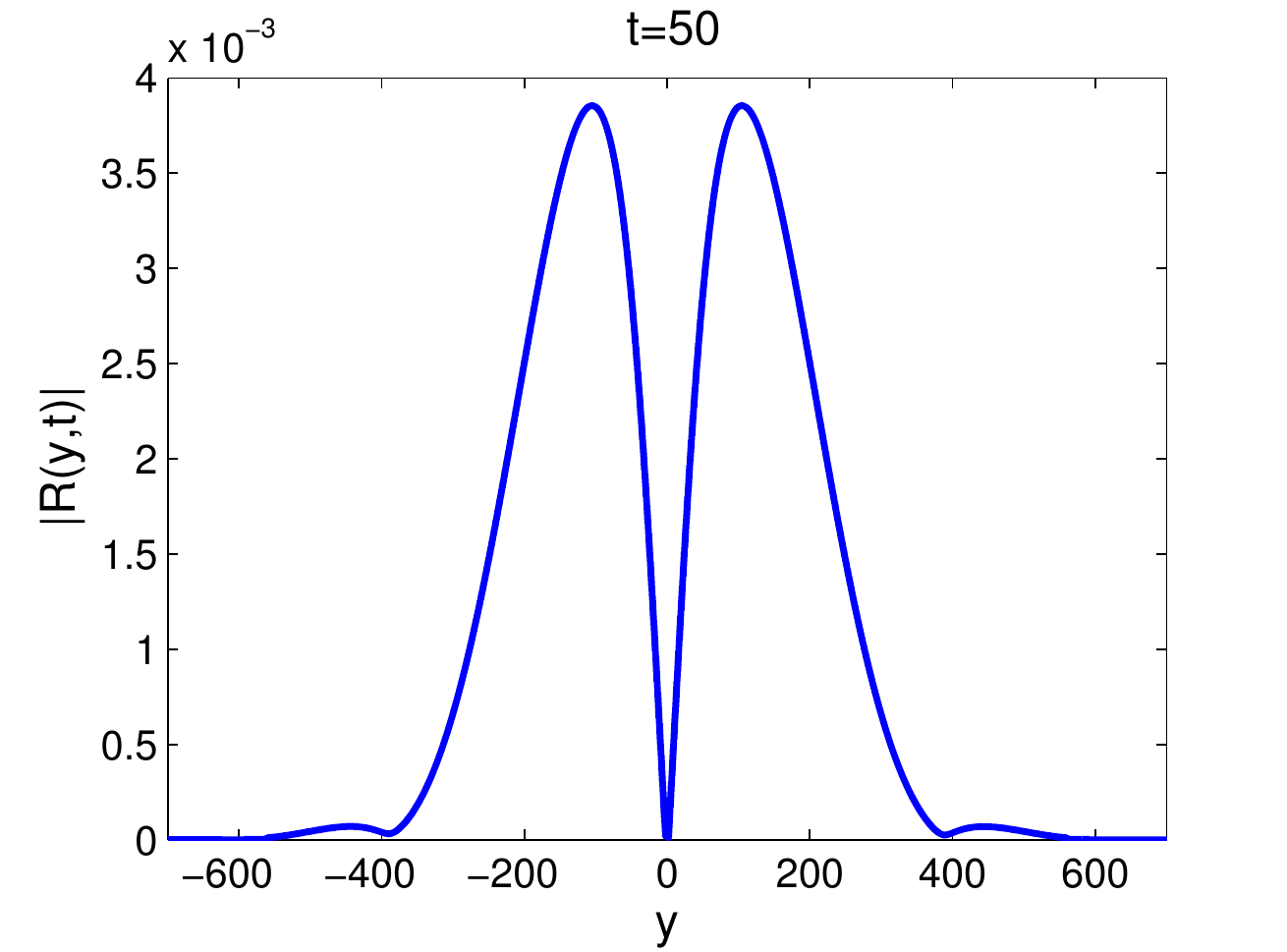,height=3.5cm,width=6cm}\\
\psfig{figure=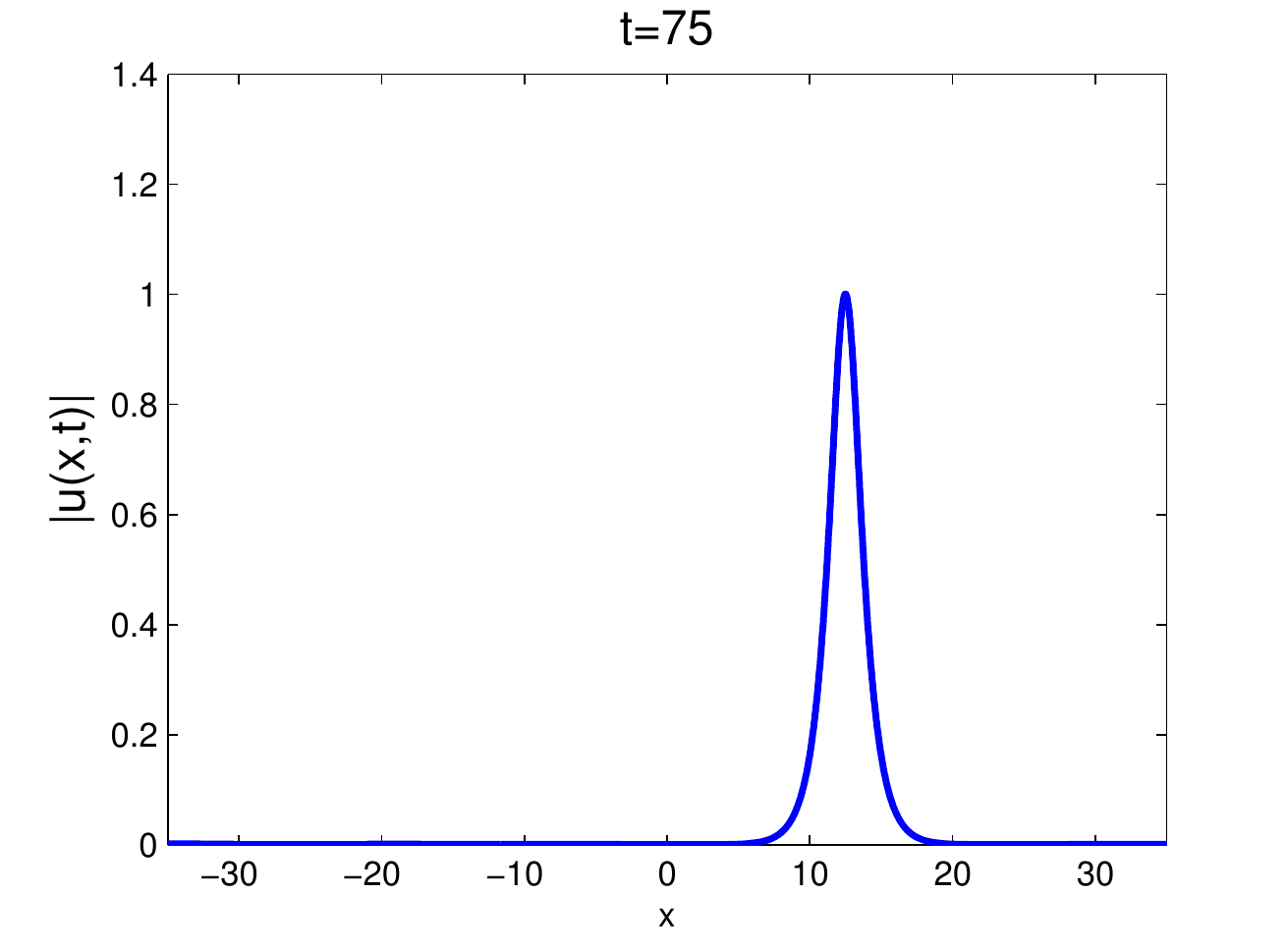,height=3.5cm,width=6cm}&\psfig{figure=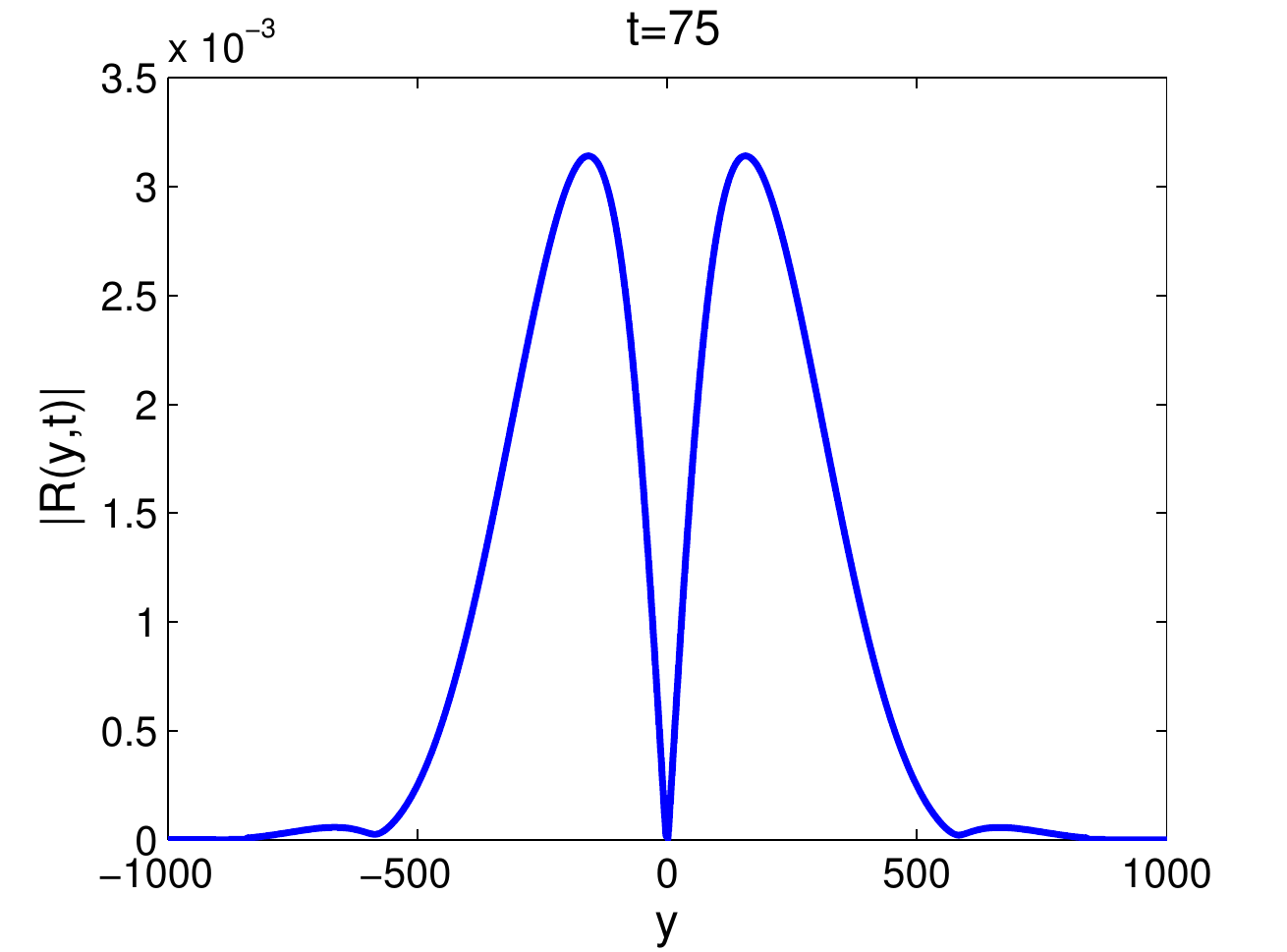,height=3.5cm,width=6cm}\\
\psfig{figure=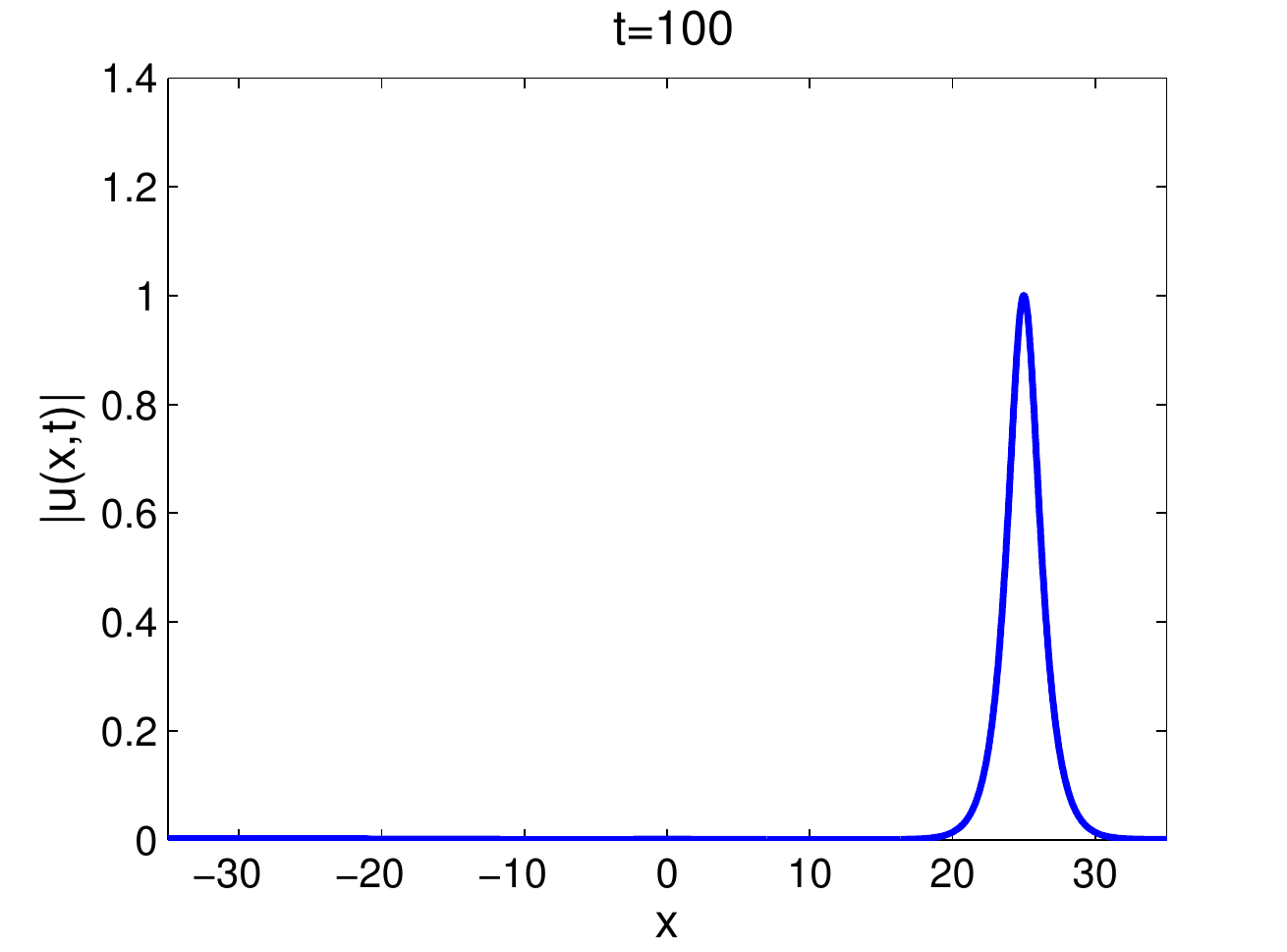,height=3.5cm,width=6cm}&\psfig{figure=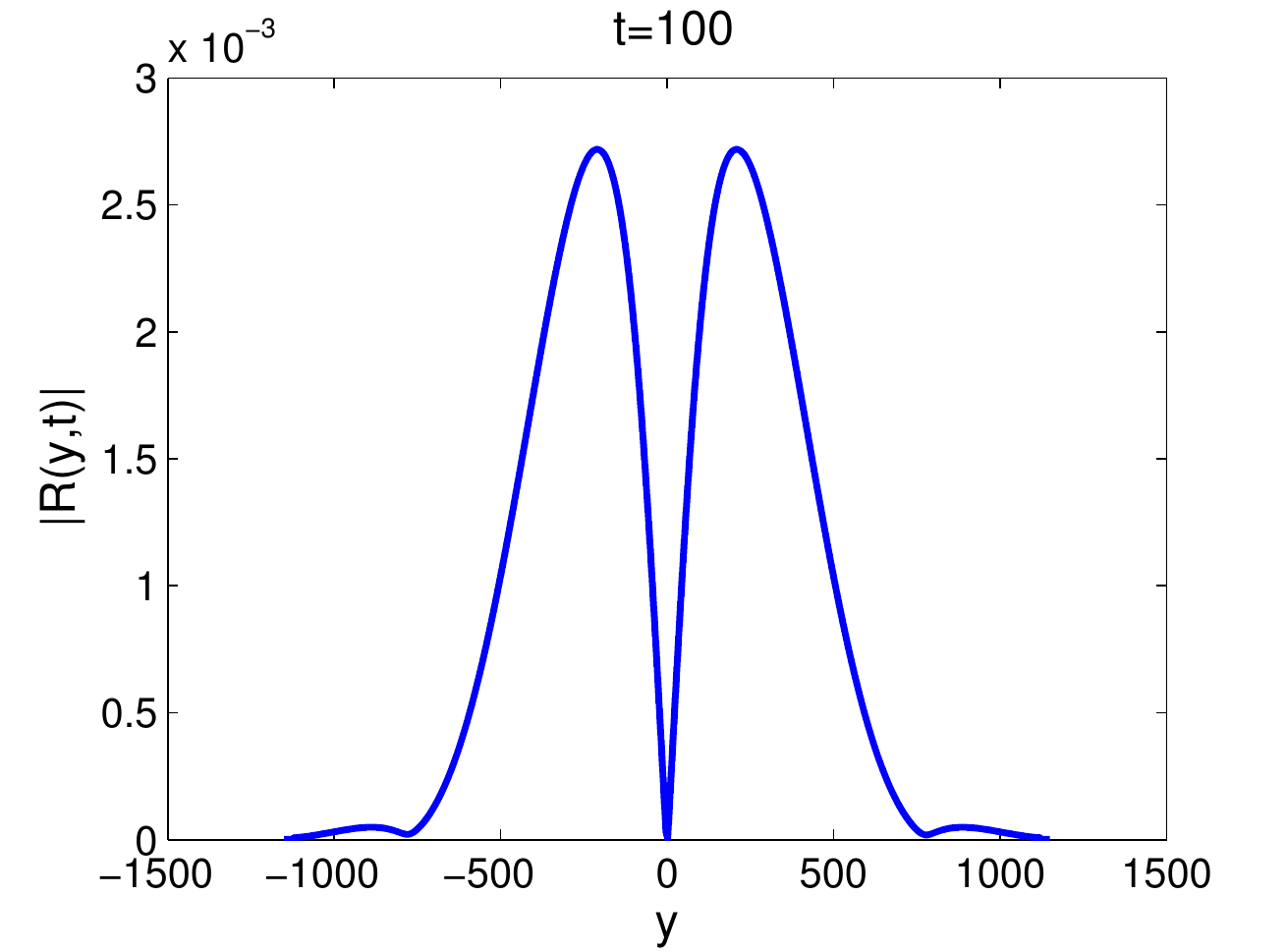,height=3.5cm,width=6cm}
\end{array}
$$
}
\caption{Dynamics of $|u(x,t)|$ and $|R(y,t)|$ with $v_0=0.5$ and $D_0=-25$ till $t=100$. }\label{fig:longtime}
\end{figure}

With $D_0=-25$ the long time dynamics of the multichannel solution in the numerical example (\ref{cubic})-(\ref{R ini}) on a large domain $\Omega=[-1150,1150]$ is shown in Fig. \ref{fig:longtime}.

%\begin{figure}[t!]
%{$$
%\begin{array}{cc}
%\includegraphics[height=3.5cm,width=5.5cm]{uv05t25-eps-converted-to.pdf}&\includegraphics[height=3.5cm,width=5.5cm]{Rv05t25-eps-converted-to.pdf}\\
%\includegraphics[height=3.5cm,width=5.5cm]{uv05t50-eps-converted-to.pdf}&\includegraphics[height=3.5cm,width=5.5cm]{Rv05t50-eps-converted-to.pdf}\\
%\includegraphics[height=3.5cm,width=5.5cm]{uv05t75-eps-converted-to.pdf}&\includegraphics[height=3.5cm,width=5.5cm]{Rv05t75-eps-converted-to.pdf}\\
%\includegraphics[height=3.5cm,width=5.5cm]{uv05t100-eps-converted-to.pdf}&\includegraphics[height=3.5cm,width=5.5cm]{Rv05t100-eps-converted-to.pdf}
%\end{array}
%$$
%}
%\caption{Dynamics of $|u(x,t)|$ and $|R(y,t)|$ with $v_0=0.5$ and $D_0=-25$ till $t=100$. }\label{fig:longtime}
%\end{figure}

Based on Figs. \ref{fig2}-\ref{fig3}, we can draw the
following observations:
\begin{enumerate}
\item The collective coordinates $v(t)$, $D(t)$, $\omega(t)$ and $\gamma(t)$ converge the steady state as $t$ goes to infinity (cf. Fig. \ref{fig2}), and the dispersive part $R(y,t)$ spreads out to far field (cf. Fig. \ref{fig3}).

\item The dynamics of each component of $\sigma(t)$ is not monotone in time $t$ (cf. the left figure in Fig. \ref{fig2}). This indicates that the process of the dynamics of the soliton and the dispersive wave is not monotone.

\item The dispersive wave $\phi$ in this case has a large expanding velocity (cf. Fig \ref{fig3}). It is because that the chosen initial perturbation, i.e. the $R_0$ in (\ref{R ini}), has a large $H^1$ norm. For the kind of situation, we remark that using the absorbing boundary conditions could be a more efficient way of study.
\end{enumerate}

%\begin{figure}[t!]
%{$$
%\begin{array}{cc}
%\includegraphics[height=7.0cm,width=5.5cm]{2dphi0.pdf}&\includegraphics[height=7.0cm,width=5.5cm]{2dR0.pdf}\vspace{-3.5cm}\\
%\includegraphics[height=7.0cm,width=5.5cm]{2dphi05.pdf}&\includegraphics[height=7.0cm,width=5.5cm]{2dR05.pdf}\vspace{-3.5cm}\\
%\includegraphics[height=7.0cm,width=5.5cm]{2dphi1.pdf}&\includegraphics[height=7.0cm,width=5.5cm]{2dR1.pdf}\vspace{-2cm}
%\end{array}
%$$}
%\caption{Profiles of $\phi_\omega(\bx,t)$ and $R(\bx,t)$ at different $t$ with camera fixed in domain $[-8,8]\times[-8,8]$. }\label{fig4}
%\end{figure}

\begin{figure}[t!]
{$$
\begin{array}{cc}
\psfig{figure=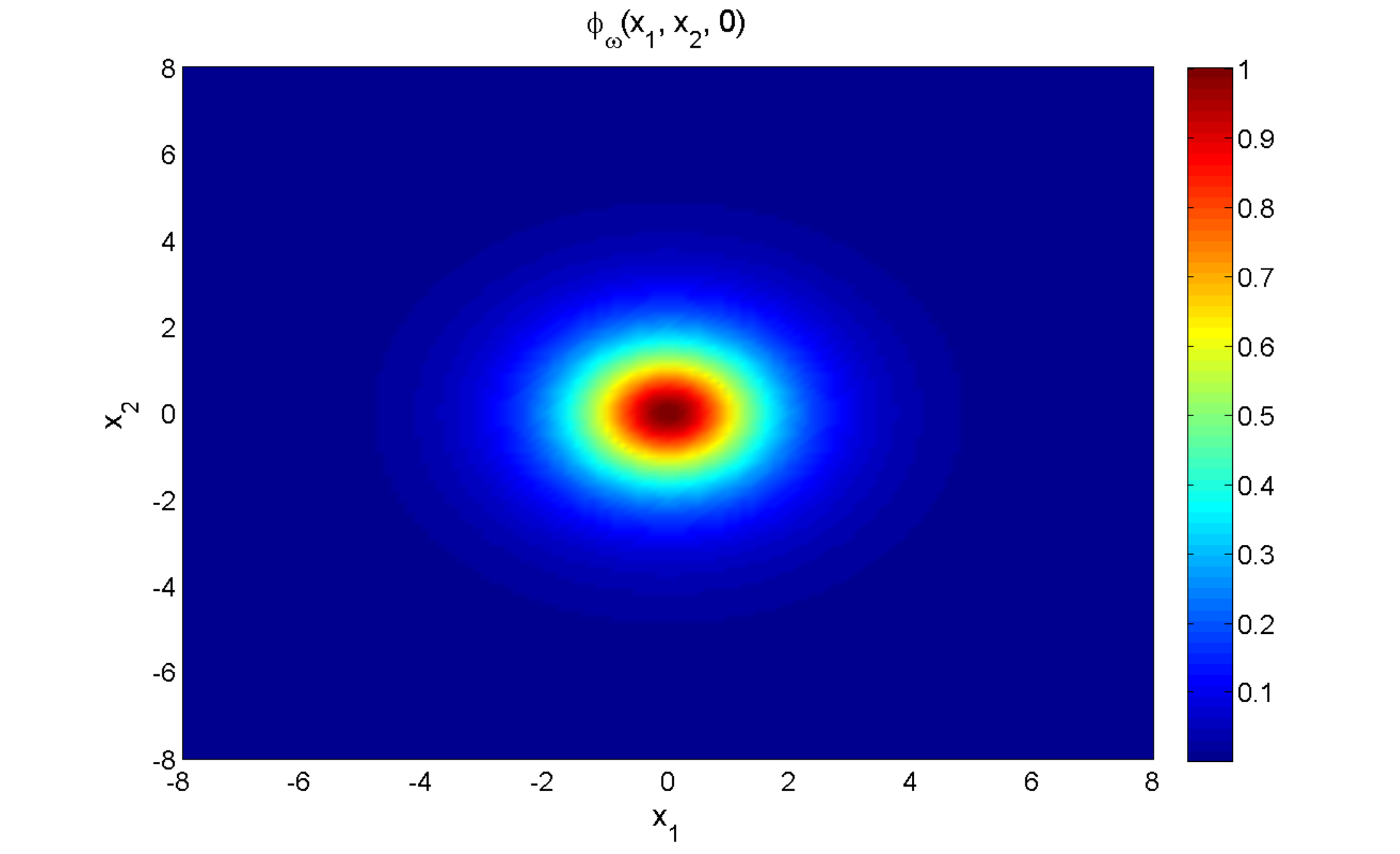,height=5.0cm,width=6cm}&\psfig{figure=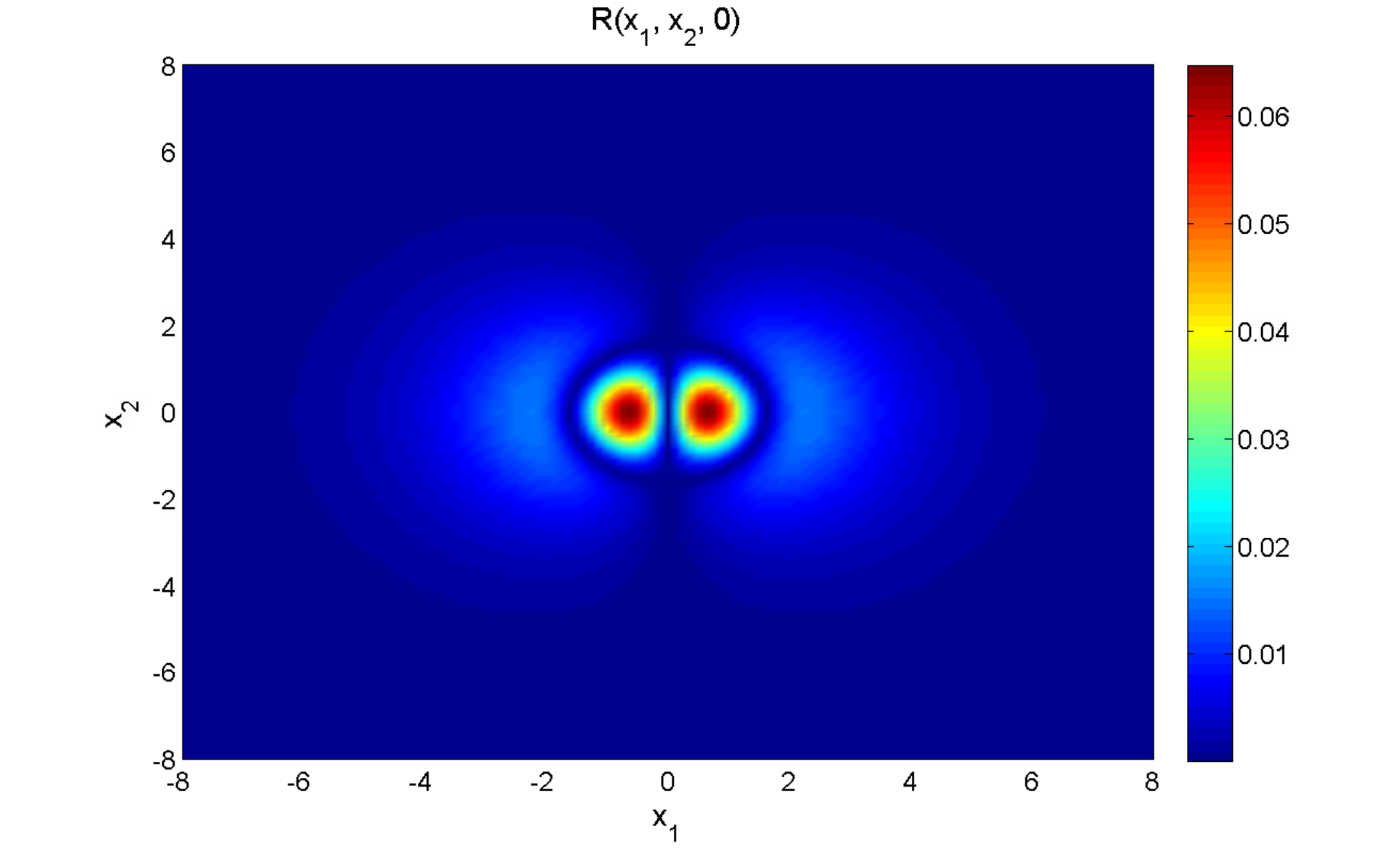,height=5.0cm,width=6cm}\\
\psfig{figure=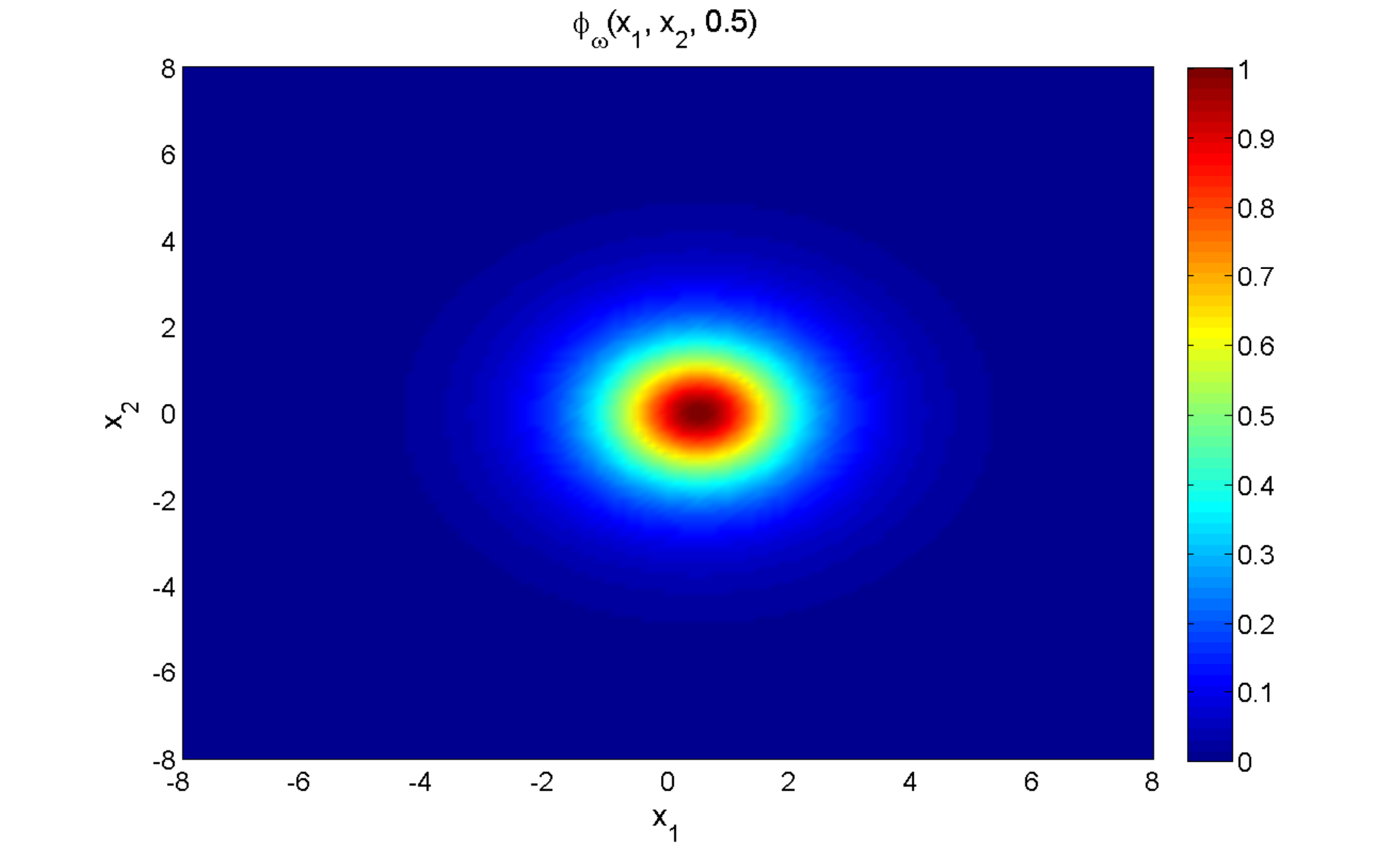,height=5.0cm,width=6cm}&\psfig{figure=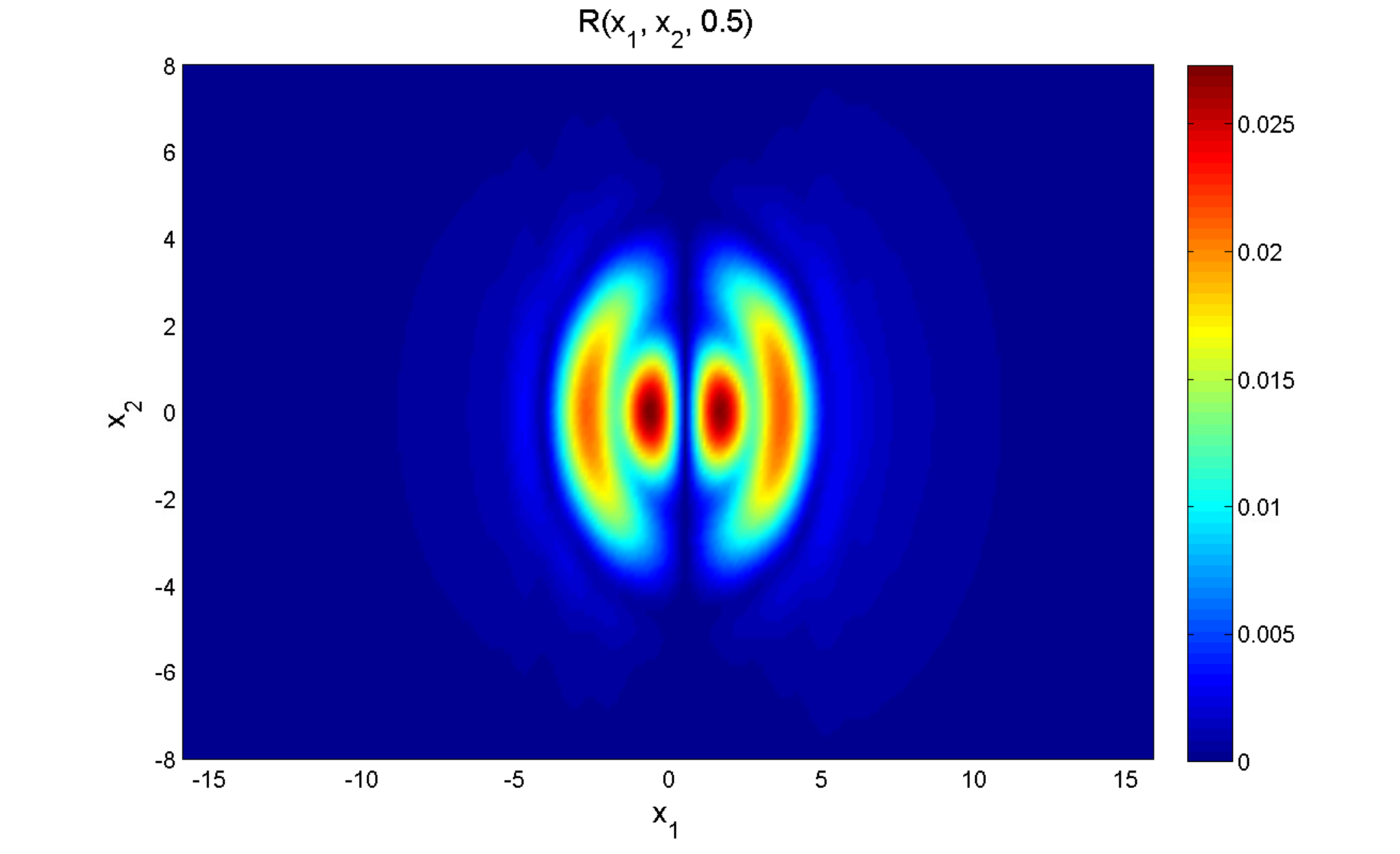,height=5.0cm,width=6cm}\\
\psfig{figure=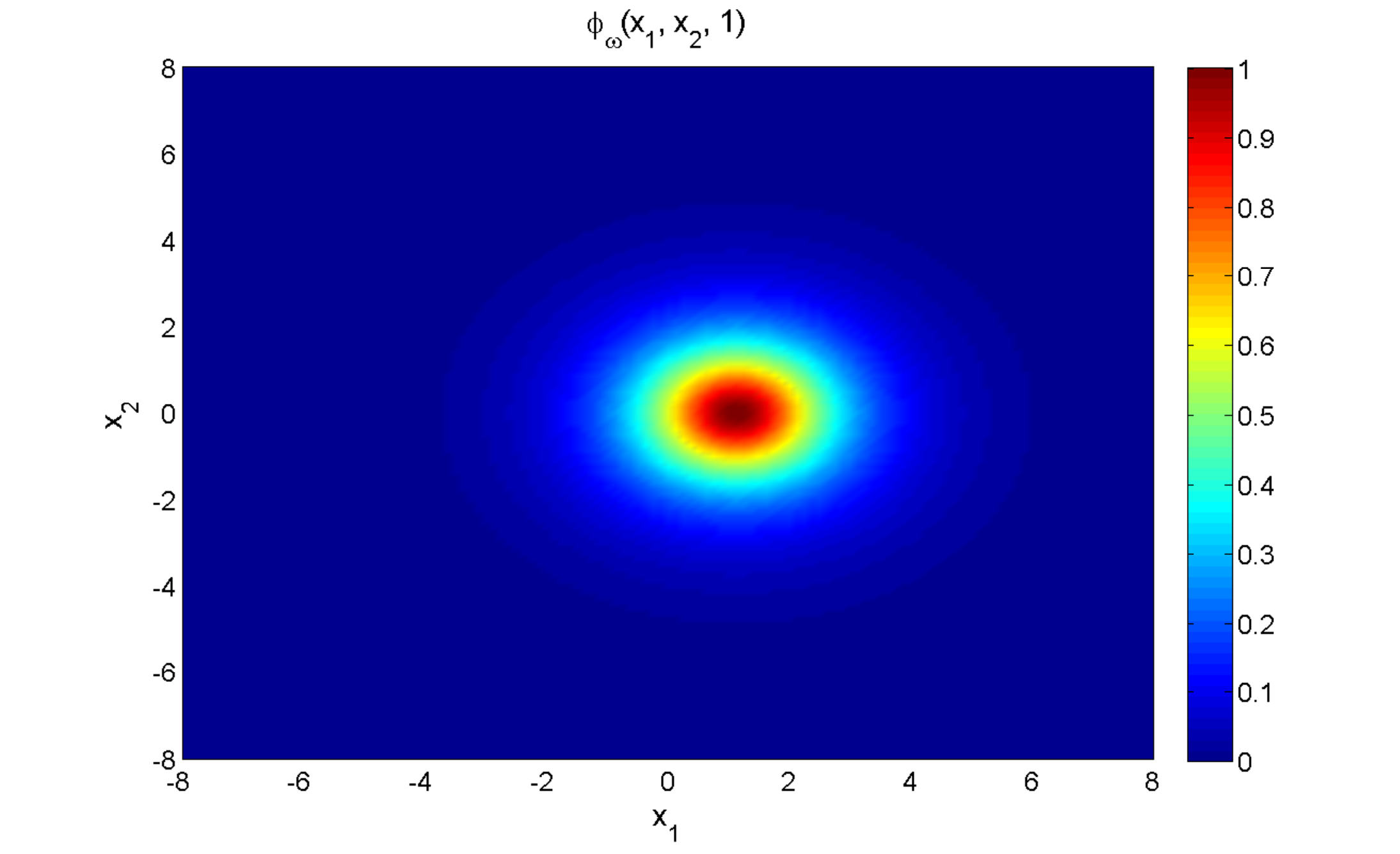,height=5.0cm,width=6cm}&\psfig{figure=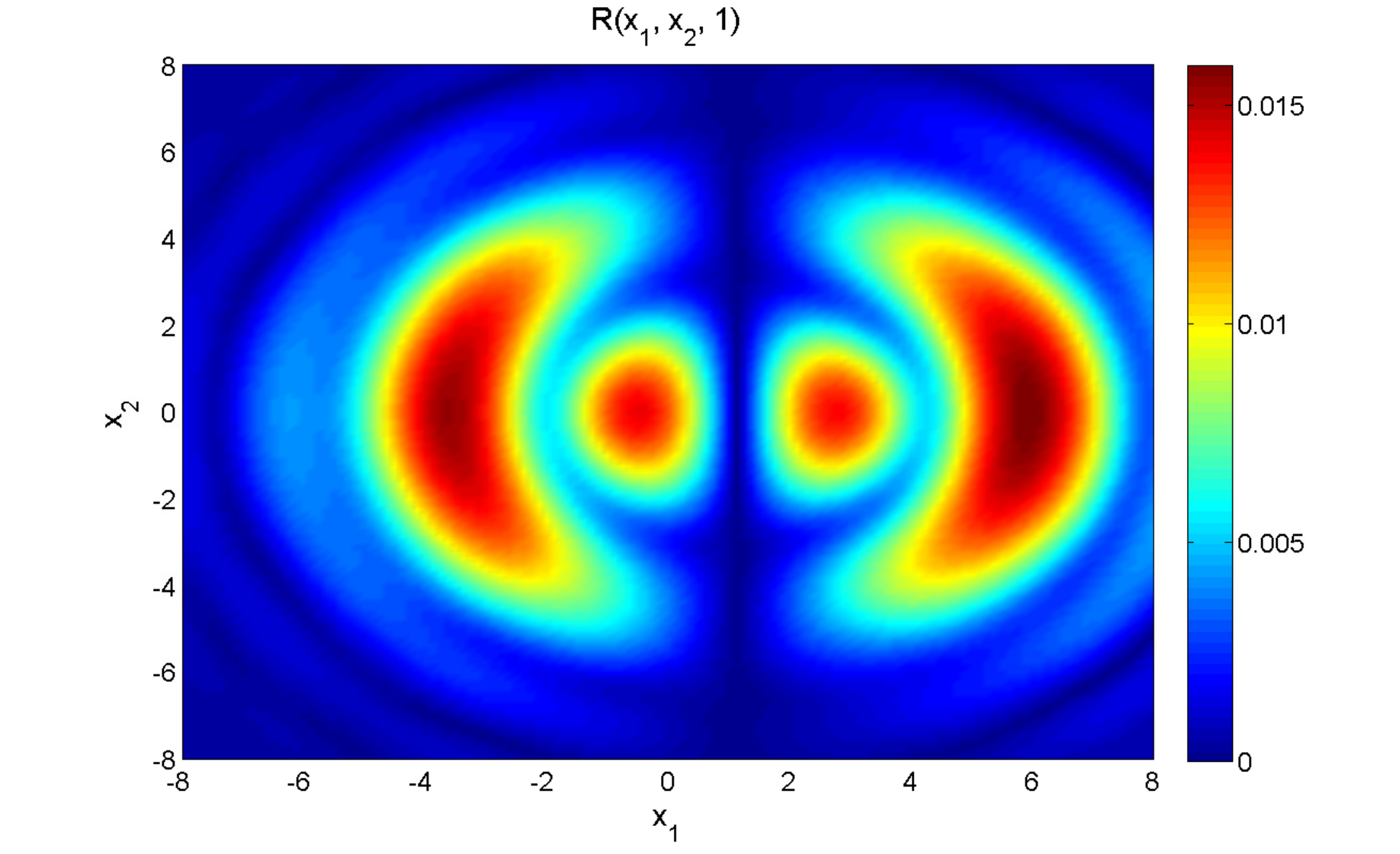,height=5.0cm,width=6cm}
\end{array}
$$}
\caption{Profiles of $\phi_\omega(\bx,t):=\phi_{\omega(t)}(\by(\bx,t),t)$ and $R(\bx,t):=|R(\by(\bx,t),t)|$ at different $t$ with camera fixed in domain $[-8,8]\times[-8,8]$. }\label{fig4}
\end{figure}

The modulation equations method also works well in two dimensions. Here we give a 2D numerical example. We take $d=2$ and
 $$\bx=(x_1,x_2),\quad \by=(y_1,y_2),\quad \bv=(v_1,v_2),\quad D=(D_1,D_2),$$
 in (\ref{nls}) and (\ref{modulation full}) with cubic nonlinearity (\ref{cubic}). We choose the initial data as
\begin{align*}
\bv_0=(1,0),\quad D_0=(0,0),\quad \gamma_0=0.5,\quad \omega_0=1,
\end{align*}
and
$$R_1(\bx,0)=0.2x_1\fe^{-x_1^2-x_2^2}-\frac{<0.2 x_1\fe^{-x_1^2-x_2^2},x_1\phi_{\omega_0}>}{\|x_1\phi_{\omega_0}\|_{L^2}^2} x_1\phi_{\omega_0},\quad R_2(\bx,0)=0.$$
and the computational domain as $\Omega=[-16,16]\times [-16,16]$. The dynamics of the soliton $\phi_\omega(\bx,t):=\phi_{\omega(t)}(\by(\bx,t),t)$ and the radiation $R(\bx,t):=|R(\by(\bx,t),t)|$ by the modulation equations are shown in Fig. \ref{fig4} with camera fixed in domain $[-8,8]\times[-8,8]$.

\section*{Acknowledgements}
The authors would like to thank the referees for their constructive comments and suggestions that greatly improved the paper.
Part of this work was done when the authors were visiting the School of Mathematics and Statistics, Central China Normal University, China, 2015.

\bibliographystyle{model1-num-names}
\bibliography{<your-bib-database>}

\end{document}